\documentclass[final,3p,times]{elsarticle}
\RequirePackage{lineno} 
\usepackage{epsfig}
\usepackage{color}
\begin{document}
\setlength{\unitlength}{1cm}
\begin{frontmatter}
\title{A Fast Algorithm for Muon Track Reconstruction and its Application to the ANTARES Neutrino
Telescope}
\author[IFIC]{J.A. Aguilar}
\author[CPPM]{I. Al Samarai}
\author[Colmar]{A. Albert}
\author[Barcelona]{M. Andr\'e}
\author[Genova]{M. Anghinolfi}
\author[Erlangen]{G. Anton}
\author[IRFU/SEDI]{S. Anvar}
\author[UPV]{M. Ardid}
\author[NIKHEF]{A.C. Assis Jesus}
\author[NIKHEF]{T.~Astraatmadja\footnote{\scriptsize{Also at University of Leiden, the Netherlands}}}
% ...
\address[IFIC]{\scriptsize{IFIC - Instituto de F\'isica Corpuscular, Edificios Investigaci\'on de Paterna, CSIC - Universitat de Val\`encia, Apdo. de Correos 22085, 46071 Valencia, Spain}}
\address[CPPM]{\scriptsize{CPPM - Centre de Physique des Particules de Marseille, Aix-Marseille Universit\'e, CNRS/IN2P3, 163 Avenue de Luminy, Case 902, 13288 Marseille Cedex 9, France}}
\address[Colmar]{\scriptsize{GRPHE - Institut universitaire de technologie de Colmar, 34 rue du Grillenbreit BP 50568 - 68008 Colmar, France }}
\address[Barcelona]{\scriptsize{Technical University of Catalonia,Laboratory of Applied
Bioacoustics,Rambla Exposici\'o,08800 Vilanova i la Geltr\'u,Barcelona, Spain}}
\address[Genova]{\scriptsize{INFN - Sezione di Genova, Via Dodecaneso 33, 16146 Genova, Italy}}
\address[Erlangen]{\scriptsize{Friedrich-Alexander-Universit\"{a}t Erlangen-N\"{u}rnberg, Erlangen Centre for Astroparticle Physics, Erwin-Rommel-Str. 1, 91058 Erlangen, Germany}}
\address[IRFU/SEDI]{\scriptsize{Direction des Sciences de la Mati\`ere - Institut de recherche sur les lois fondamentales de l'Univers - Service d'Electronique des D\'etecteurs et d'Informatique, CEA Saclay, 91191 Gif-sur-Yvette Cedex, France}}
\address[UPV]{\scriptsize{Institut d'Investigaci\'o per a la Gesti\'o Integrada de Zones Costaneres (IGIC) - Universitat Polit\`ecnica de Val\`encia. C/  Paranimf 1. , 46730 Gandia, Spain.}}
\address[NIKHEF]{\scriptsize{Nikhef, Science Park, Amsterdam, The Netherlands}}
% ...
\address[IRFU/SPP]{\scriptsize{Direction des Sciences de la Mati\`ere - Institut de recherche sur les lois fondamentales de l'Univers - Service de Physique des Particules, CEA Saclay, 91191 Gif-sur-Yvette Cedex, France}}
% ...
\address[Genova-UNI]{\scriptsize{Dipartimento di Fisica dell'Universit\`a, Via Dodecaneso 33, 16146 Genova, Italy}}

\date{\today}
\begin{abstract}

An algorithm is presented, that provides a fast and robust reconstruction
 of neutrino induced upward-going muons and a discrimination 
 of these events from  downward-going atmospheric muon background 
 in data collected by the ANTARES neutrino telescope. 
 The algorithm consists of a hit merging and hit selection procedure followed by 
 fitting steps for a track hypothesis and a point-like light source.
 It is particularly well-suited for real
 time applications such as online monitoring and fast triggering of optical 
 follow-up observations for multi-messenger studies.
 The performance of the algorithm is evaluated with Monte Carlo simulations
and various distributions are compared with that obtained in ANTARES data.

\end{abstract}

\end{frontmatter}

\section{Introduction}

The main goal of neutrino telescope experiments such as AMANDA~\cite{amanda},
IceCube~\cite{icecube}, NT-200 in lake Baikal~\cite{baikal} and
ANTARES~\cite{antares} is the observation of high energy neutrinos from non-terrestrial
sources. These instruments detect Cherenkov light from the passage of relativistic charged particles 
produced in neutrino interactions in the detector or its surrounding material. 
By measuring the photon arrival times
at known positions, the particle trajectory can be reconstructed.
The flight direction of
these particles is nearly colinear with the incident neutrino direction 
for neutrinos above 10~TeV.
The best angular resolution can be reached for $\nu_\mu, \bar\nu_\mu$ charged current interactions where the measured
particle is a muon that can travel several kilometres in water at TeV energies.
Events induced by neutrinos from astrophysical sources
must be distinguished from background
which originates in the Earth's atmosphere. 
Whereas atmospheric neutrinos are
considered a non-reducible background (at least without the use of their energy), 
atmospheric muons can in principle be excluded by simple
geometrical considerations. To exclude the contribution from downward-going atmospheric
muons, it is sufficient to identify upward-going events, 
which can only be produced by neutrinos.
This distinction requires a reliable
determination of the elevation angle, because downward-going atmospheric muons outnumber
upward-going atmospheric neutrinos by 5 to 6 orders of magnitude for typical
neutrino telescope installation depths. 

In the following, a tracking algorithm is presented which meets this goal.
Apart from providing an excellent up-down separation, it is very fast 
and therefore well-qualified for real time applications. This has been demonstrated 
in ANTARES, where the algorithm can cope with a trigger rate of 100~Hz, 
running on a single PC. It is used in various online monitoring tasks, 
for an analysis of atmospheric muons~\cite{atmmuon} and 
as an alert sending program to trigger optical follow-ups of selected neutrino 
events~\cite{tatoo}. Other potential applications include the study of
magnetic monopoles and nuclearites and the measurement of atmospheric neutrinos.
The presented algorithm is complementary to other reconstruction methods which have
been developped in ANTARES~\cite{ronald,aart}.

Section 2 describes the ANTARES detector, Section 3 explains the
geometrical approximations which are used in all subsequent steps. The hit
merging and hit selection methods are detailed in Sections 4 and 5 whereas the
fitting procedure can be found in Section 6. The performance of the entire
algorithm introduced in Sections 2-6 is discussed in Section 7.

\section{The ANTARES detector}

The ANTARES detector is located about 40 km off the coast of Toulon, France, at a depth
of 2475 m in the Mediterranean Sea ($42^\circ 48'$N, $6^\circ 10'$E).
It consists of 12 flexible lines, each
with a total height of 450 m, separated by distances of 60-70 m
from each other. They are anchored to the sea bed and kept near
vertical by buoys at the top of the lines. Each line carries a total of
75 10-inch Hamamatsu photo multipliers (PMTs) housed in glass spheres, 
the optical modules (OM)~\cite{om}, arranged in 25 storeys (three optical modules per
storey) separated by 14.5 m, starting 100~m above the sea floor.
Each PMT is oriented $45^\circ$ downward with respect to the vertical.
A titanium cylinder in each storey houses the electronics for readout
and control.

The positions of the line anchors on the sea bed and the construction details 
(cable length {\it etc.}) of each detector line 
are well known and stable in time.
However the absolute positions and orientation of the 
individual OMs vary with time as the lines move freely in the sea current. 
Their movements are monitored by a system of acoustic transponders
and receivers distributed over the lines and on the sea bed. After each cycle of acoustic signal
exchange between these elements, a triangulation provides the shape of each line. This is 
complemented by data from a compass card within each electronics cylinder which measures 
the heading and the inclination of its storey. 
The processing of these combined data provides the position and orientation of each OM 
with a precision of better than 10 cm, but after an important delay.

\section{Geometrical approximations}

To qualify for real time applications the presented method uses only time independent geometrical
informations and the following approximations are applied: 
\begin{description}
\item[Line shape] Detector lines are considered to be perfectly vertical. 
Line distortions due to sea currents are ignored.
\item[Storey geometry] The detailed geometry of the storey is ignored. 
Each storey is considered as a single OM, which is located 
directly on the detector line with an axis-symmetric field of view.
The signals of the three PMTs within one storey are combined.
\end{description}
These approximations are used in the hit merging, hit selection and fitting steps.
Their effect on the angular resolution is discussed in Section~\ref{sec:results}.
Under the above approximations, a detector line is a straight line in space.
It is also assumed that a muon track is a straight line, {\it i.e.} multiple scattering {\it etc.}
is ignored.
Except for the special case that the muon track and a detector line are parallel 
(in the case of an exactly vertical track), the point of closest approach 
of the muon track to the detector line can be determined. Most of 
the Cherenkov light is expected to be seen in the vicinity of this point. 
This is used in the hit selection as well as 
in the fitting procedure of the algorithm described below.

\section{Hits and hit merging}

The PMT signals are digitized by a custom
built ASIC chip~\cite{ars}. If the analog PMT signal crosses a
preset threshold, typically 1/3 photoelectron (p.e.), its arrival time is 
measured together with its charge. The latter is obtained by integrating the
analog PMT signal within a time window of 40~ns. Each such pair of time and charge is called a
hit and the corresponding data for each hit are sent to shore. The data
stream is processed by a computer farm in the shore station which
searches for different physics signals according to predefined trigger
conditions. If at least one trigger pattern is found, all hits within a time window of 
few $\mu$s around the trigger time are stored on disk. The DAQ system is described in 
detail in~\cite{daq}.

For the data which will be presented in Section~\ref{sec:results} two trigger
algorithms are used. Both rely on {\it L1 hits} which are
defined either as 2 hits in coincidence within 20~ns in two OMs
on the same storey or as a single hit with an amplitude
larger than a threshold of 3 or 10~p.e..
One of the used triggers requires at least five causally connected L1 hits with respect to a
given track direction. The second used trigger requires two pairs of L1 hits in adjacent 
or next-to-adjacent storeys within 80~ns and 160~ns respectively. 

Before starting the actual hit selection, time and charge calibrations are applied to 
transform raw data to physical units such as time (in nanoseconds) and 
charge (in equivalent number of photoelectrons).
Up to now the standard data handling in
ANTARES has been described. The following steps are instead related 
to the presented algorithms. Different reconstruction methods might apply 
other hit treatment and hit selection procedures.

The calibrated hits are time-ordered for each storey. 
All hits on the same storey within a predefined time window are merged, 
as applied to early ANTARES data in~\cite{ronald}.
Merging is performed by adding the hit charges and keeping the time of the earliest hit.
The merging time window has to be chosen short enough to ensure a clean hit selection,
with the earliest of the merged hits having a high probability of originating 
from a particle track.
On the other hand, the merging time window must be long enough 
to select most of the signal hits. 
For a 10-inch phototube, operated at a threshold of 0.3~photoelectrons
a typical noise rate of 60-100~kHz is observed in sea water~\cite{biolum}. 
In this case 
a merging window of 20~ns is appropriate. 
As a result, a list of merged hits per storey is obtained
and is the basis for all subsequent steps.

\section{Hit selection}
\label{sec:hit-selection}
\begin{figure}
\centering
\centering
    \begin{picture}(14,7)
       \epsfclipon 
       \epsfxsize=7cm
       \put(0,0){\epsfbox{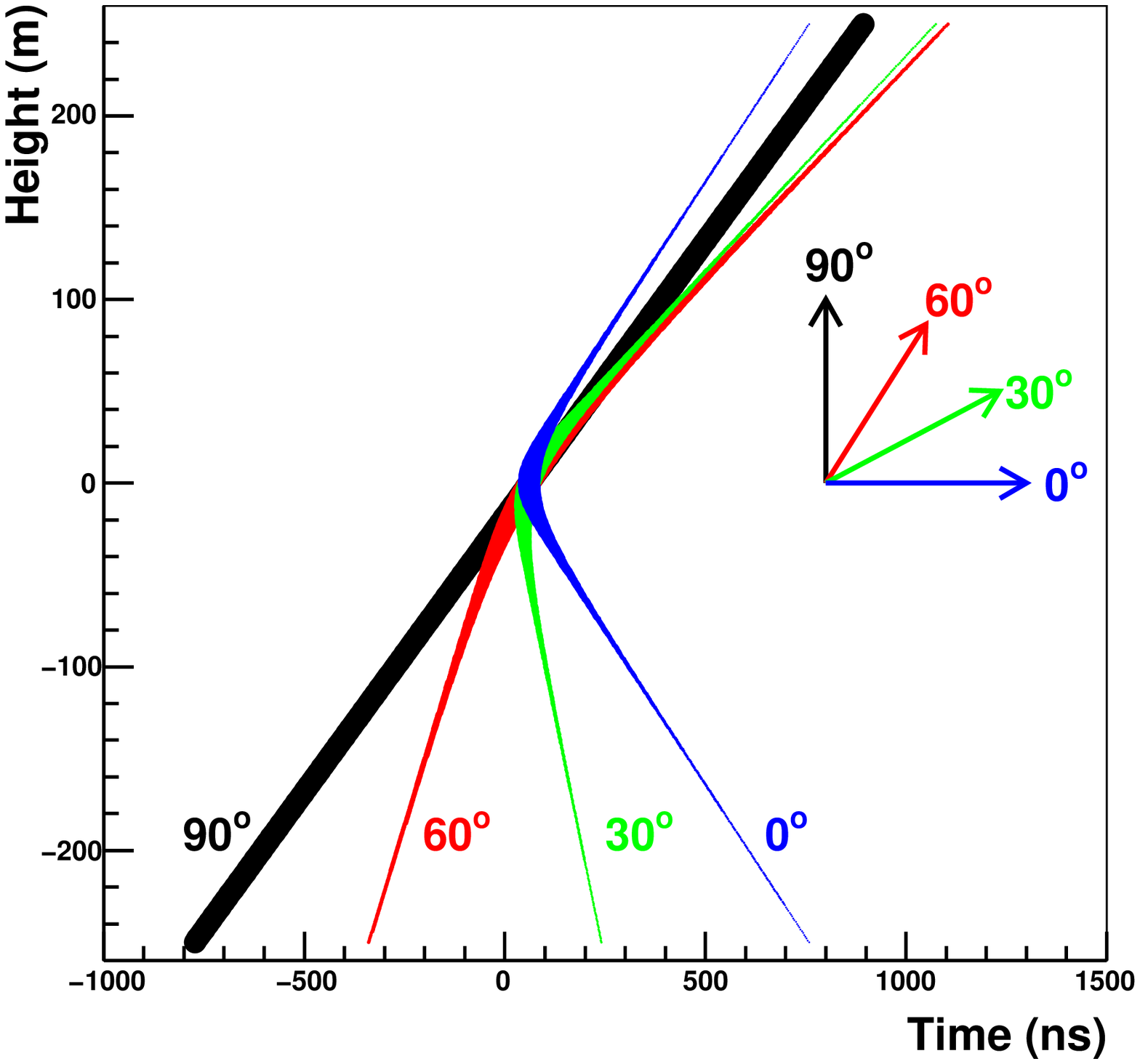}}
       \put(7,0){\epsfbox{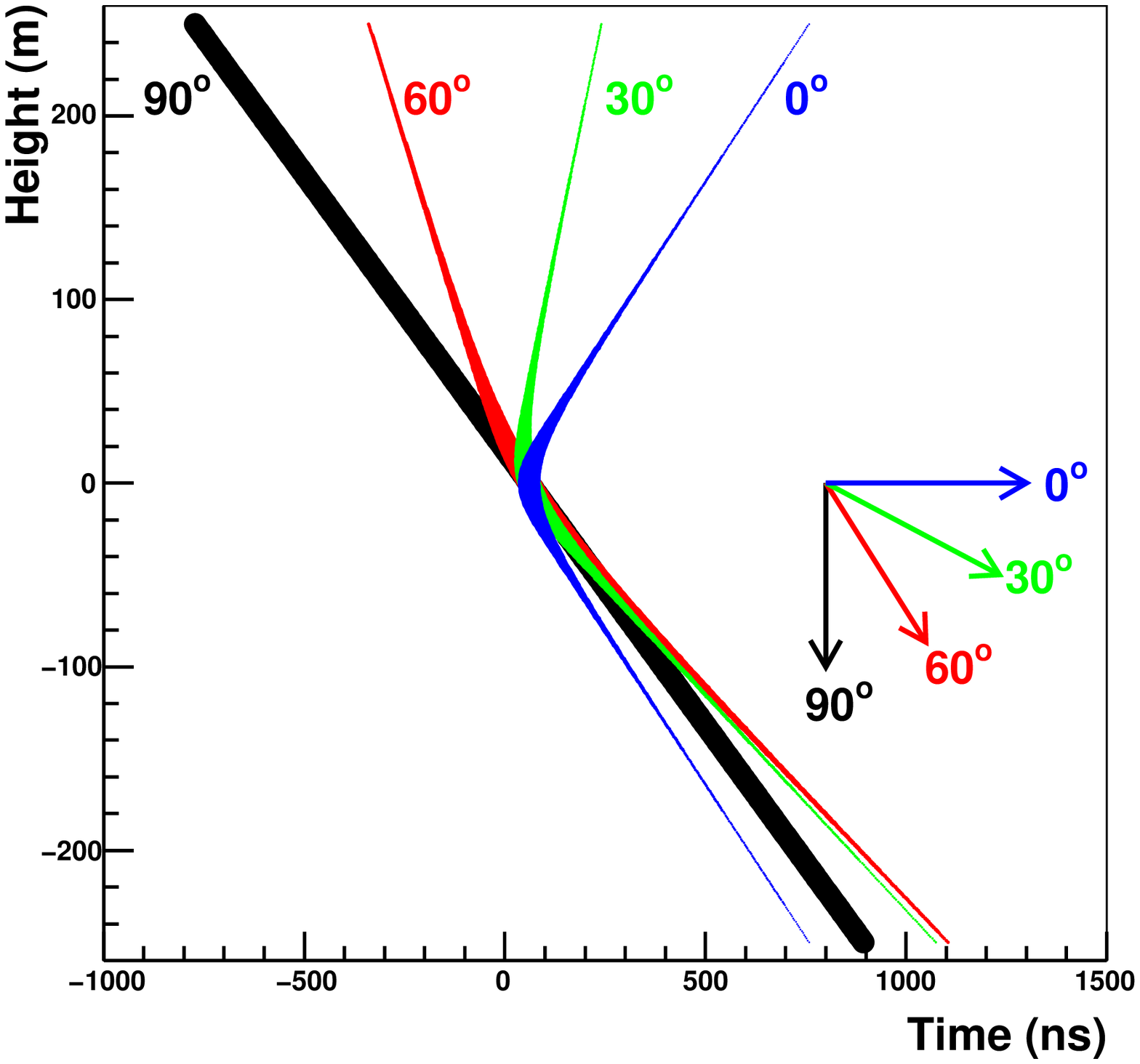}}
    \end{picture}
\caption{The pattern of Cherenkov light from a muon track; ({\bf left}: upward-going; 
{\bf right}: downward-going) on a single detector line. 
Vertical position (z-coordinate) along the detector line and photon arrival time 
are related by a hyperbola
(see Equations~\ref{thit},\ref{xgam}). The point 
$z=0,t=0$ defines the point of closest approach
between track and detector line; here, the track passes at 10~m. The colours correspond to different
track angles: black $90^\circ$ (vertical), red $60^\circ$, green $30^\circ$, blue $0^\circ$ (horizontal).
The thickness of the lines indicates the brightness of the expected signal.}
\label{fig:z-t-scheme}
\end{figure}

The light pattern expected from a highly relativistic charged particle is illustrated 
in Figure~\ref{fig:z-t-scheme}. 
The goal of the hit selection is to choose hits due to Cherenkov photons 
and avoid random hits from the optical background or scattered late hits.
Late hits due to scattering are rare compared to typical optical background 
of 60-100~kHz.
The probability of a Cherenkov photon at $\lambda=470$~nm to suffer from light scattering 
on a path of 60~m (the inter-line distance and coincidentally the water absorption length) 
is only few per cent~\cite{light}.
 
For tracks which are not parallel to a detector line, 
most of the Cherenkov light is expected near the point of closest
approach of the track to the detector line. Therefore, the first step of the hit selection 
consists of finding a 
``hot spot" of light on a detector line which originates from Cherenkov light 
induced by a passing
particle. 
In the presence of background light, a single high charge hit does 
not provide a ``hot spot" with sufficient purity; two high charge hits in adjacent or 
next-to-adjacent storeys are needed.
The detector storeys are numbered 
consecutively along the z-axis, like floors in a building. 
For a given floor $i$ the adjacent or next-to-adjacent storeys
are at floor $i\pm j$ with $j=1,2$.   
To ensure that hits on floors $i$ and $i\pm j$ 
were caused by the same particle, the following condition on the absolute hit time 
difference $\Delta t$ is imposed
\begin{equation}
       \Delta t < j \Delta z \frac{n}{c} + t_s,
\label{eq:eq0}
\end{equation} 
with $\Delta z$ the absolute vertical distance between adjacent storeys, 
$c$ the speed of light in vacuum, $n$ the refractive index of the medium and $t_s$ an additional
time delay which accounts for the timing uncertainty of the photon detection itself
and the simplifications of the detector geometry. It is assumed that storeys 
are equally spaced.
Such a causality condition has been
used for hit selection prior to track reconstruction in Baikal~\cite{baikal} and
AMANDA~\cite{amanda-reco}.
For the ANTARES detector with $\Delta z = 14.5~$m and $t_s=10~$ns, one finds $\Delta t < 80~$ns
for adjacent and $\Delta t < 150~$ns for next-to-adjacent storeys. 
Only merged hits which originate from different PMTs or with a charge above 2.5~photoelectrons 
are considered here, because they have a high probability of being caused by a particle track.
All hits which contribute to the above defined ``hot spots" are stored as selected hits for the next steps.
If several ``hot spots" occur in the same storey at different times, only the earliest of them
is kept.

For the following steps, only detector lines having at least one ``hot spot" are considered. 
This condition removes all isolated hits from the sample.
Using the ``hot spot" hits as seeds, hits are added to the list of selected hits. 
From this point on, the hit selection is exclusively based on timing information.
No charge cut is applied at this stage.
This allows to use the attenuated signal
at larger distances from the point of closest approach as illustrated in Figure~\ref{fig:z-t-scheme}.
New hits on a given detector line are recursively sought within a narrow time window around the times 
of already selected hits. 
The expected hit times in 
adjacent or next-to-adjacent storeys are calculated, 
assuming the hits arrive linearly in the $z$-$t$ plane as indicated by the asymptotes of the hyperbola
in Figure~\ref{fig:z-t-scheme}. 
The projected arrival time $t_{i\pm j}$ at floor $i\pm j, (j=1,2)$ is related to any pair out of the three hit
times $t_i,t_{i\mp 1},t_{i\mp 2}$ at floors $i,i\mp 1,i\mp 2$ in the following way
 \begin{eqnarray}
     t_{i\pm j} &=& t_i + j \left( t_i-t_{i\mp 1} \right), \\
     t_{i\pm j} &=& t_i + j \left( t_i-t_{i\mp 2} \right)/2, \\     
     t_{i\pm j} &=& t_{i\mp 1} + (j+1) \left( t_{i\mp 1}-t_{i\mp 2} \right). 
\label{eq:eq1}
\end{eqnarray}
In the following, time windows $[t_{i\pm j}^{early},t_{i\pm j}^{late}]$ based on the above equations 
are defined, which are used for the hit selection.
If only two out of the three hits in floors $i,i\mp 1,i\mp 2$ exist, only one of the 
extrapolations above can be performed. 
A new hit in floor $i\pm j$ is accepted if it occurred within a time window of
 \begin{eqnarray}
        t_{i\pm j}^{early} &=& t_{i\pm j}-j t_s \\
        t_{i\pm j}^{late} &=& \mbox{max}(t_i + j \Delta z \frac{n}{c}+ t_s,t_{i\pm j}+j t_s),
\label{eq:timewindow}
\end{eqnarray}
with $t_s$ and $\Delta z$ from Equation~\ref{eq:eq0}. This
time window is asymmetric according to the generic shape of the Cherenkov cone hyperbola in the 
$z$-$t$ plane (see Figure~\ref{fig:z-t-scheme} and Equations~\ref{thit},\ref{xgam} discussed later).
If all three hits exist, the minimum and maximum $t_{i\pm j}$ are determined and a time window of
 \begin{eqnarray}
	t_{i\pm j}^{early} &=& t_{i\pm j}^{min}-j t_s \\
	t_{i\pm j}^{late} &=& t_{i\pm j}^{max}+j t_s,
\label{eq:timewindow3}
\end{eqnarray}
is used for floor $i\pm j$. 
If several hits in floor $i\pm j$
meet the criteria, only the earliest one is selected.
This procedure is applied recursively.
If new hits cannot be found in either floor $i\pm 1$ or $i\pm 2$, the procedure stops, 
avoiding gaps of more than one floor.
The occurrence of several well-separated ``hot spots" 
on the same detector line serves as independent seeds to restart the procedure.
The described hit selection is performed once. It yields at most one hit per storey.
The result of the hit selection on a bright neutrino candidate event from 2008 
can be seen in Figure~\ref{fig:evdisplay}.

\section{Fitting procedure}

Before starting any fit, the list of selected hits is examined. 
Only events with more than 5 hits are accepted.
If all selected hits are on a single detector line, a procedure 
for a single-line fit is started,
otherwise a multi-line fit procedure is performed.
Both procedures are discussed in the following.

\subsection{Fitting particle tracks}
\label{sec:trackfit}

\begin{figure}
\centering
\epsfig{figure=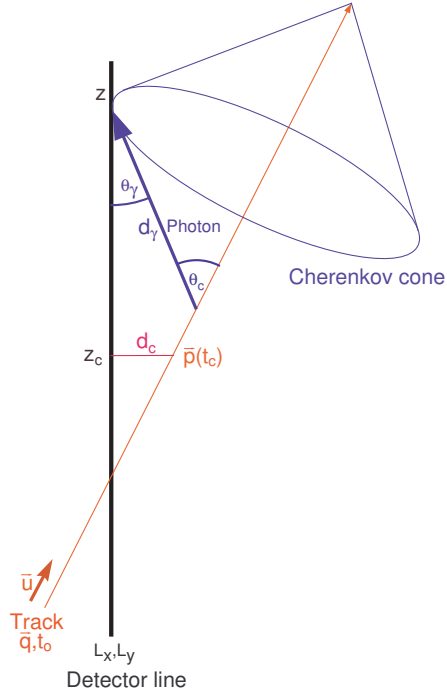,width=6.0cm}
\caption{Illustration of the variables used to describe a track and its corresponding
Cherenkov cone with respect to a vertical detector line.}
\label{fig:sketch}
\end{figure}

A particle track is considered to be a straight line in space. 
The particle is assumed to move with the speed of
light in vacuum. All space-time points, $\vec{p}(t)$, that are part of the track can
be parameterized as 
\begin{equation}
\vec{p}(t) = \vec{q} + c(t-t_0)\vec{u}.
\label{track}
\end{equation}
The particle passes through the point $\vec{q}$ at time $t_0$ (i.e. $\vec{q}=\vec{p}(t_0)$) and moves in the direction $\vec{u}$. 
The vector $\vec{q}$ can be
shifted along the track by redefining $t_0$. Therefore the track is defined by a total of 5 parameters: 
three
values to fix $\vec{q}$ for a given time and two angles to define $\vec{u}$:
\begin{equation}
\vec{u} = \left\{\cos\theta\cos\phi,\cos\theta\sin\phi,\sin\theta\right\},
\end{equation} 
with $\theta$ the elevation angle and $\phi$ the azimuth angle.
The detector lines are approximated as vertical
lines along the $z$-axis (see Section 2) at fixed horizontal positions $(L_x,L_y)$. 
The z-component of the point of closest approach of a particle track to a detector line is given by
\begin{equation}
z_c = \frac{q_z-u_z(\vec{q}\cdot\vec{u})+u_z(L_xu_x+L_yu_y)}{1-u_z^2},
\label{zcm}
\end{equation}
through which the particle passes at time
\begin{equation}
t_c = t_0 + \frac{1}{c}(L_xu_x+L_yu_y+z_cu_z-\vec{q}\cdot\vec{u}),
\label{tcm}
\end{equation}
at a distance
\begin{equation}
d_c = \sqrt{(p_x(t_c)-L_x)^2 + (p_y(t_c)-L_y)^2}.
\label{dcm}
\end{equation}
The above-defined variables are illustrated in Figure~\ref{fig:sketch}.
If the track is exactly vertical, and therefore parallel to the detector line, 
$t_c=t_0$ and $z_c=q_z$ are chosen.
The track can be conveniently reparametrized in terms of $z_c, t_c, d_c$ for each detector line
and the two angles which define $\vec{u}$. For a single-line fit the detector line can be placed 
at the coordinate origin $(L_x,L_y)=(0,0)$ and Equations~\ref{zcm},\ref{tcm},\ref{dcm} simplify to
\begin{equation}
z_c = \frac{q_z-u_z(\vec{q}\cdot\vec{u})}{1-u_z^2},
\label{zc}
\end{equation}
\begin{equation}
t_c = t_0 + \frac{q_z u_z-(\vec{q}\cdot\vec{u})}{c(1-u_z^2)},
\label{tc}
\end{equation}
\begin{equation}
d_c = \sqrt{p_x^2(t_c) + p_y^2(t_c)}.
\label{dc}
\end{equation}
A single detector line is invariant against a rotation
of the coordinate system around the z-axis. This means the track is not fully defined 
and only 4 parameters  $z_c, t_c, d_c$ and $u_z$ can be determined.
No dependence from azimuth is observed.

To
perform the track fit,
it is necessary to know (a) the arrival time $t_\gamma$ 
of a Cherenkov photon at 
the detector line positions $(L_x,L_y,z)$, (b) its corresponding travel path $d_\gamma$
and (c) its
inclination with respect to the detector line, $\cos\theta_\gamma$. 
All three values can be derived from the parameters defined above and the
refractive index $n$ which is related to the Cherenkov angle $\theta_c$ by $1/n=\cos\theta_c$:
\begin{equation}
t_\gamma(z) = (t_c-t_0) +\frac{1}{c}\left((z-z_c)u_z+\frac{n^2-1}{n}d_\gamma(z)\right),
\label{thit}
\end{equation}
\begin{equation}
d_\gamma(z) = \frac{n}{\sqrt{n^2-1}}\sqrt{d_c^2+(z-z_c)^2(1-u_z^2)},
\label{xgam}
\end{equation}
\begin{equation}
\cos\theta_\gamma(z)= (1-u_z^2)\frac{z-z_c}{d_\gamma(z)}+\frac{u_z}{n}.
\label{ctgam}
\end{equation}
These equations hold exactly for Cherenkov photons of a given wavelength. 
Dispersion and group velocity effects as
well as delays due to light scattering are ignored. 
An appropriate effective refractive index is used, depending on the medium in
which the detector is installed.

\subsection{Fitting point-like light sources: the bright-point fit}

In contrast to a particle track, a `bright-point' is defined as a point-like light source which emits 
a single light flash at a
given moment. The light emission is assumed to be isotropic. 
The model of a bright-point does not only apply to artificial light
sources such as optical beacons installed for detector calibrations~\cite{ledbeacon} 
but also in some approximation to light from hadronic
and electromagnetic showers, for which the actual extension of the shower is typically 
smaller than the detector line distances. 
A bright-point is defined by four parameters: its position $\vec{q}$ and its time $t_0$.
In analogy with the definitions of the point of closest approach for particle tracks, 
it is straightforward to see that for a bright-point $z_c=q_z$, $t_c=t_0$ and 
\begin{equation}
d_c = \sqrt{(q_x-L_x)^2 + (q_y-L_y)^2},
\label{bpdcm}
\end{equation}
which simplifies to
\begin{equation}
d_c = \sqrt{q_x^2 + q_y^2},
\label{bpdc}
\end{equation}
for a single detector line at $(L_x,L_y)=(0,0)$. In this case, only
the three parameters $z_c, t_c, d_c$ can be determined,
which again means that the number of parameters is reduced 
by one.
The photon arrival time $t_\gamma$, its travel length $d_\gamma$ 
and its angle with respect to a given arrival 
point $z$ along the detector line can thus be determined in analogy 
to the case of a particle track:
\begin{equation}
d_\gamma(z) = \sqrt{d_c^2+(z-q_z)^2},
\label{bpxgam}
\end{equation}
\begin{equation}
t_\gamma(z) = t_0 + \frac{n}{c}d_\gamma,
\label{bpthit}
\end{equation}
\begin{equation}
\cos\theta_\gamma(z)= \frac{(z-q_z)}{d_\gamma}.
\label{bpctgam}
\end{equation}
As in the treatment of particle tracks, it is assumed that all photons have the same wavelength,
which corresponds to a fixed effective refractive index.

\subsection{Quality function}

\begin{figure}[htbp]
    \centering
    \begin{picture}(16,16)
      \epsfclipon 
      \epsfxsize=16cm 
      \put(0,0){\epsfbox{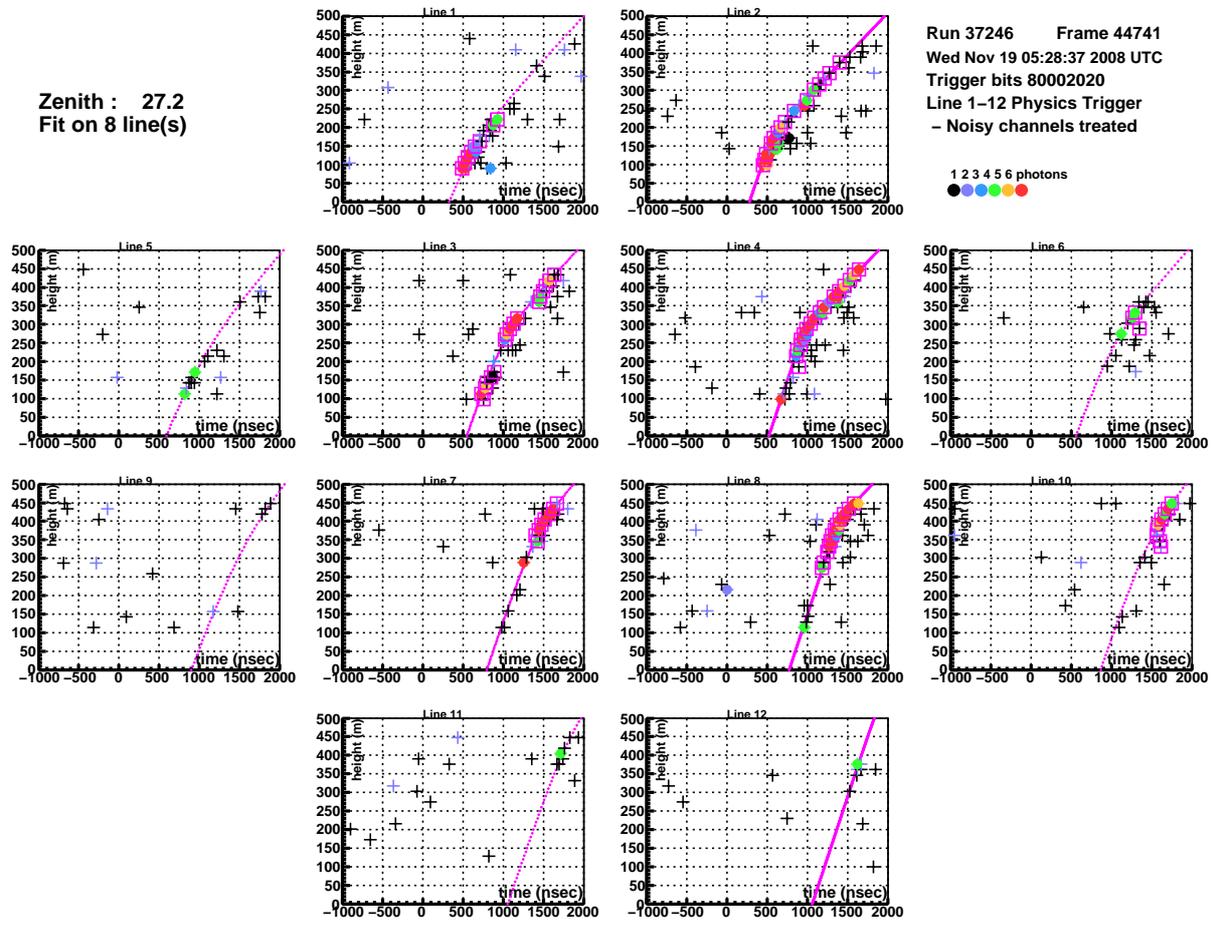}}
    \end{picture}
\caption{Event display of a bright neutrino event from the 2008 data taking period. The arrangement
of the 12 histograms corresponds approximately to the octagonal layout of the 12 detector lines 
on the sea floor. For each line the vertical position above the sea floor (m) is given as function of
the hit time (ns) (see also Figure~\ref{fig:z-t-scheme}). $t=0$ corresponds to the time of the first hit which participated in the trigger. A
window of $-1000/+2000$~ns is shown with respect to this time. $z=0$ indicates the sea floor. 
Active detector elements are between $z=100$ and $z=450$~m. Circles denote hits which participated in the trigger,
crosses are other hits, boxes indicate the hit selection of Section~\ref{sec:hit-selection}. Colors
refer to the hit charge as shown in the legend on the top right part of the figure. The line width and
style of the fitted track illustrates the minimal distance between track and detector line;
thick and solid lines stand for closer distances than thin or dotted lines.}
\label{fig:evdisplay}
\end{figure}

The quality function is based on the time differences between the hit times $t_i$ 
and the expected arrival time $t_\gamma$ of
photons from the track or bright-point, as in a standard $\chi^2$ fit. 
The quality function is extended 
with a term that accounts for measured hit 
charges $a_i$ and the calculated photon travel distances $d_\gamma$. 
The full quality function is
\begin{equation}
Q = \sum^{N_{hit}}_{i=1}\left[\frac{(t_\gamma-t_i)^2}{\sigma^2_i}+\frac{A(a_i)D(d_\gamma)}{\langle a \rangle d_0}\right],
\label{chi2}
\end{equation}
where $\langle a \rangle$ is the average hit charge calculated from all hits which have
been selected for the fit. The normalisation term $d_0$ and the functions $A(a_i)$ and $D(d_\gamma)$
are discussed below.
The second term of the right hand side of equation~\ref{chi2} exploits the fact that 
an accumulation of storeys with high charges (hot spots) 
is expected on each detector line at its point of closest approach to the track. 
If such hot spots on several detector lines are arranged in a way that
their z,~t coordinates indicate an upward-going pattern, 
the event has indeed a high probability to originate from an upward-going neutrino. 
This allows the isolation of a high purity neutrino sample. The concept is illustrated in 
Figure~\ref{fig:evdisplay} where the arrangement of the ``hot spots'' on the various detector lines allows
to classify this event as upward-going even without an overlaid fitted track.
The quality function depends
on a number of parameters which have been tuned for sea water~\cite{biolum,light} and the 
ANTARES detector~\cite{antares}.

The timing uncertainty, $\sigma_i$ is set to 10~ns for $a_i > 2.5$~photoelectrons and to 
20~ns otherwise. These values are significantly larger than the intrinsic 
transit time spread of the ANTARES phototubes (1.3~ns). They take into account an additional
smearing due to the applied geometrical approximations from Section 3 
and due to the occasional presence of late hits from small angle scattering 
or from electromagnetic processes which passed the hit selection.
 
The charge term of the quality function is not
arranged as a difference between expected value and measured charge 
in order to avoid incorrect penalties from large expected
charges. Instead, the chosen form gives a heuristic penalty to the combination of high charge 
and large distance. The
product is divided by the average charge of all the selected hits in an event, 
$\langle a \rangle$, to compensate for the fact that more energetic tracks or
showers will produce more light at the same distance. The normalisation $d_0$
balances the weight between the two terms. For ANTARES a value of $d_0=50$~m has been chosen, 
motivated by
the fact that at this distance the typical signal in a photodetection unit 
which points straight into the
Cherenkov light front is of the order of one photoelectron. 

The charges have to be
corrected for the angular acceptance of the storey. Assuming a downward-looking 
hemispherical photodetection surface with a homogeneous efficiency for photodetection
(a good approximation for an ANTARES storey), the following
simple correction function for the hit charges $a_i'$ is obtained:
\begin{equation}
a_i'=\frac{2 a_i}{\cos\theta_\gamma + 1}.
\label{wang}
\end{equation}
The average charge $\langle a \rangle$ is calculated from these corrected hit charges:
\begin{equation}
\langle a \rangle=\frac{1}{N_{hit}}\sum_1^{N_{hit}}a_i'.
\label{totamp}
\end{equation}
Furthermore, charges are protected against extreme values by the following function:
\begin{equation}
A(a_i)=\frac{a_0 a_i'}{\sqrt{a_0^2+a_i'^2}}.
\label{amp}
\end{equation}
The function $A(a_i)$ introduces an artificial saturation such that
for $a_i' \ll a_0$, the charge is relatively unaffected, {\it i.e.}
$A(a_i) \approx a_i'$, whereas for $a_i' \gg a_0$, 
the charge saturates at $A(a_i) \approx a_0$, the chosen saturation value. 

The photon travel distance is similarly protected using the function
\begin{equation}
D(d_\gamma)=\sqrt{d_1^2+d_\gamma^2},
\label{dist}
\end{equation}
which introduces a minimal distance $d_1$.
For large distances, $d_\gamma \gg d_1$, $D(d_\gamma) \approx d_\gamma$, whereas for very small
distances, $d_\gamma \ll d_1$,  $D(d_\gamma) \approx d_1$. 
This avoids an excessive pull of the fitted trajectory towards the detector line. 

Equations~\ref{amp} and \ref{dist} can be motivated by the following consideration.
The intensity of Cherenkov light decreases linearly with distance 
(ignoring absorption, dispersion and similar effects). Thus one
can write $a_i'd_\gamma=a_0d_1$. For ANTARES, the constant $a_0d_1$ should be around 
50~m$\times$photoelectron, which corresponds to the observation 
that about 50 photoelectrons can be measured for
a minimum ionizing particle at 1~m distance, or equivalently 1 photoelectron 
is seen at a distance of 50~m. Using this
identity, it follows that $A(a_i)D(d_\gamma)=a_i'd_\gamma$, {\it i.e.} 
the two functions have no effect on the product of charge times distance for Cherenkov
light; however, for background light they produce the desired penalty.

\subsection{Minimization procedure}
\label{minuit}
The MIGRAD function of the MINUIT package~\cite{minuit} is used to 
determine the minimum of the quality function Q.
The value of $Q/NDF$ (with $NDF$ the number of degrees of freedom of the fit) 
after minimisation is called {\it fit quality} $\tilde{Q}$ 
and is used for further selections.
For the single-line bright-point fit, the parameters determined are 
the point of closest approach  ({\it i.e.} $z_c, t_c, d_c$) 
with an additional parameter $u_z$ in the case of a track fit.
Equations~\ref{thit}-\ref{ctgam} and \ref{bpxgam}-\ref{bpctgam}
are used to obtain $d_\gamma,t_\gamma$ and $\cos\theta_\gamma$ for particle 
tracks and bright-point fits, respectively.

The multi-line fits use the same set of equations. But in this case one more parameter is needed to 
evaluate them. For the multi-line track fit $\vec{q}, u_z$ and $\phi$ are used, for the
multi-line bright-point fit $\vec{q}$ and $t_c$.
Table~\ref{tab:param} summarizes the number of fit parameters for the different cases.
\begin{table}[htpb]
\begin{center}
\begin{tabular}{||c|c|c||} \hline
        & single-line & multi-line \\ \hline
particle track & 4 &  5 \\
bright-point & 3 & 4 \\ \hline
\end{tabular}
\end{center}
\caption{Number of parameters for the different fits. Since multi-line events do not 
have rotational symmetry, an additional parameter is required.}
\label{tab:param}
\end{table}

A critical issue is the choice of the starting values and possible boundary conditions.
As explained above, the set of fit parameters used for single-line bright-point fits 
is $z_c,t_c,d_c$. 
Considering
that most of the hits are grouped around the point of closest approach,
$z_c\approx\langle z_i \rangle$ at $t_c\approx\langle t_i \rangle$, the mean values 
of the selected hit positions and times are used as starting values. 
The initial choice of $d_c$ is not critical 
for the fit stability. It is set to $d_c=10$~m. This completely determines 
the starting
values for the single-line bright-point fit. For the single-line track-fit,
a starting value for $u_z$ has to be chosen as well.
To avoid finding a wrong $u_z$, due to the inherent ambiguities when fitting Cherenkov light cones, 
three different starting values of $u_z$ are tried: $-0.9, 0$ and $0.9$. 
The fit with the smallest fit quality $\tilde{Q}$ is taken.
The allowed parameter range for $u_z$ is limited to physical values ({\it i.e.} $-1 \le u_z \le 1$).
Fit results for which the fit stops at a boundary are excluded. 

The multi-line bright-point fit starts from the center of gravity of the selected hits in 
space and time.
The multi-line track fit requires, however, a prefit. 
For this purpose, an improved version~\cite{aart} of the ``DUMAND prefit"~\cite{dumand} is used, 
which fits a straight line to the hit positions while allowing for a variable effective 
particle speed.

\section{Test of the algorithm with Monte Carlo simulations and ANTARES data}
\label{sec:mcsims}

The entire algorithm as described in Sections 3-6
was tested with ANTARES data and simulations of atmospheric muons and neutrinos.

\subsection{Data sample}

For the following analysis data are used which have been taken from December 2007 until end of 2008.
During this time a minimum of 9 detector lines was active. From May 2008 on the ANTARES detector
was complete with its 12 lines. Only runs are used, which have at least 80\% of active channels 
when averaged over the run. 
A channel is called active if it has an instantaneous counting rate of at least 40~kHz, 
well below the typical baseline noise rate of 60~kHz. 
Further, runs are excluded for which more than 40\% of the active channels have
been in {\it burst mode}, which means they measure an instantaneous counting rate more than
20\% above the usual baseline rate. This selection provides a data sample which corresponds to an
active livetime of 173~days. 

\subsection{Simulations}

Downward-going atmospheric muons were simulated with Corsika~\cite{corsika}. 
The primary flux was composed of several nuclei according to~\cite{nsu} and the 
QGSJET hadronic model~\cite{qgsjet} was used in the shower development.
An average Mediterranean atmosphere is simulated, 
{\it i.e.} one which corresponds to a situation typically found in spring or autumn. 
Seasonal variations of the muon flux due to atmospheric temperature and density variations 
are found to be smaller than 3\% in this region~\cite{season}, and they are ignored in this
analysis. 
Upward-going neutrinos were simulated according to the parameterization of 
the atmospheric $\nu_\mu$ flux from~\cite{bartol} in the energy range from 10~GeV to 10~PeV. 
Only charged current interactions of $\nu_\mu$ and $\bar\nu_\mu$ were considered.
Disappearance of muon neutrinos
due to neutrino oscillations was included in a simplified two-flavor model assuming 
maximal mixing and $\Delta m^2 = 2.4\cdot 10^{-3}$eV$^2$.

The Cherenkov light, produced in the vicinity of the detector, was propagated taking into account
light absorption and scattering in sea water~\cite{light}.
The angular acceptance, quantum efficiency and other characteristics of the PMTs
were taken from~\cite{om} and the overall geometry corresponded to the layout of the
ANTARES detector~\cite{antares}. The optical noise has been variable during the considered data taking
period. It is simulated from counting rates observed in real data. At the same time the definition of
active and inactive channels has been applied from real data runs of the selected period.
The generated statistics corresponds to an equivalent observation time of 10 years for atmospheric
neutrinos and, depending on primary cosmic ray energy, 
from 2 weeks up to 1 year for atmospheric muons.

A dedicated simulation has been done to study the impact of coincident atmospheric muon events.
One month of equivalent livetime has been simulated using the MUPAGE code~\cite{mupage}. 
An increase of 0.1\% is found at trigger level. The fraction of downward reconstructed tracks
increases by a similar fraction whereas for misreconstructed upward-going tracks an increase of 
0.2\% is found without using a cut in the fit quality. 

\subsection{Results}
\label{sec:results}

\begin{figure}[htbp]
    \begin{picture}(16,5)
      \epsfclipon 
      \epsfxsize=7cm 
      \put(0,0){\epsfbox{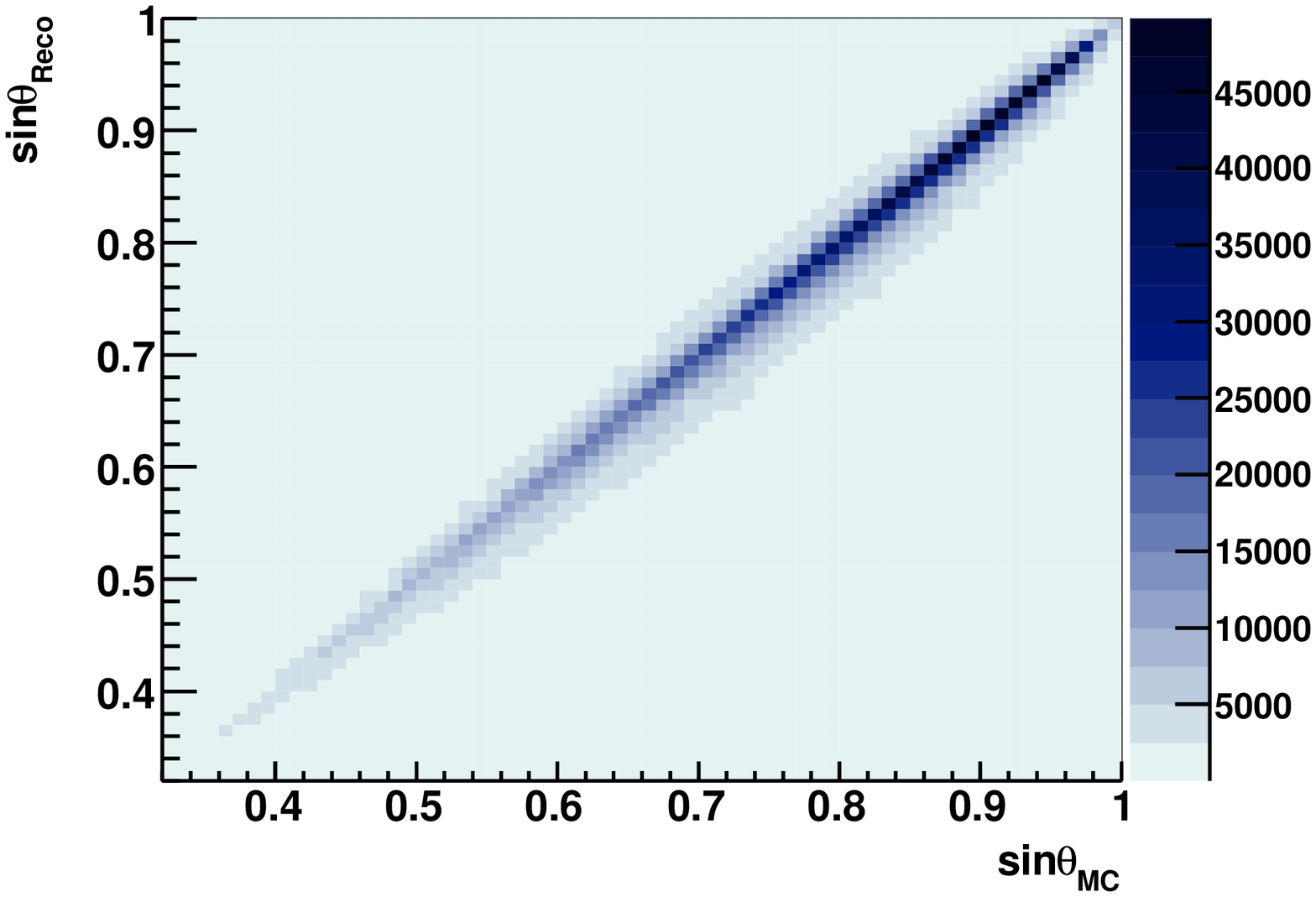}}
      \put(8,0){\epsfbox{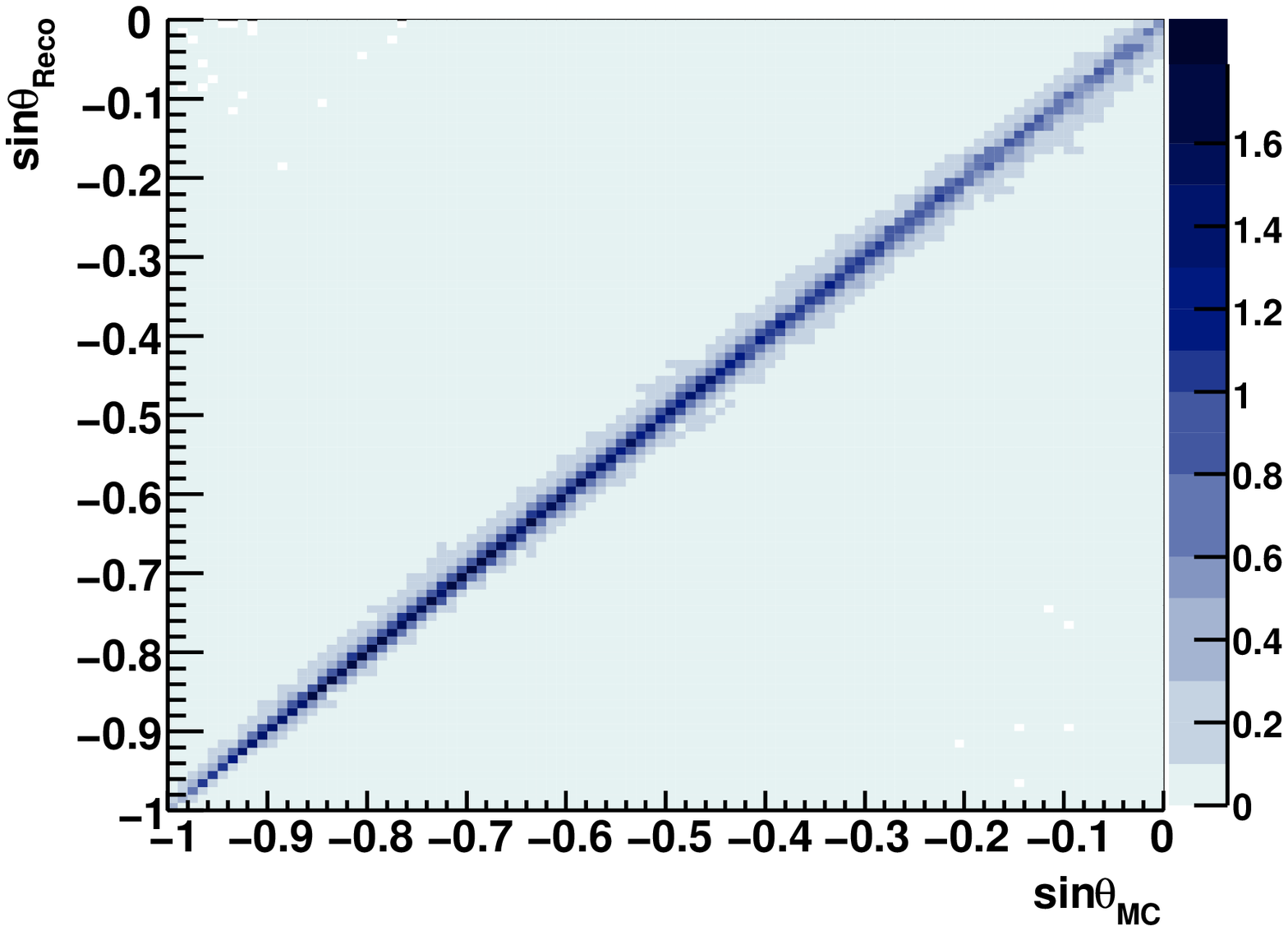}}
    \end{picture}
\caption{Correlation between reconstructed and simulated elevation angle;
{\bf Left}: down-going atmospheric muon events; {\bf Right}: upward-going atmospheric neutrino events.}
\label{fig:mu-corr}
\end{figure}

The accuracy in the reconstruction of the elevation angle 
(the angle with respect to the horizontal plane),
which needs to be good for
up-down separation of the reconstructed tracks, is illustrated in 
Figure~\ref{fig:mu-corr}. The reconstructed elevation angle is plotted versus
the true elevation angle for down-going atmospheric muons and upward-going atmospheric neutrinos, 
that are reconstructed as multi-line events. No further event selection was applied. 
More than 95\% of the fits converge for
all events with selected hits on at least two lines. 
Most events are located within a narrow band around the 
diagonal and 80\% of the events have their elevation angle reconstructed to better
than $5^\circ$. No other structures are visible on the plots. It should be noted that this 
excellent agreement is obtained, even though more than half of the triggered
atmospheric muon events are muon bundles, 
whereas here just a single track reconstruction is performed.

\begin{figure}
\centering
    \begin{picture}(16,6)
      \epsfclipon 
      \epsfxsize=8cm 
      \put(0,0){\epsfbox{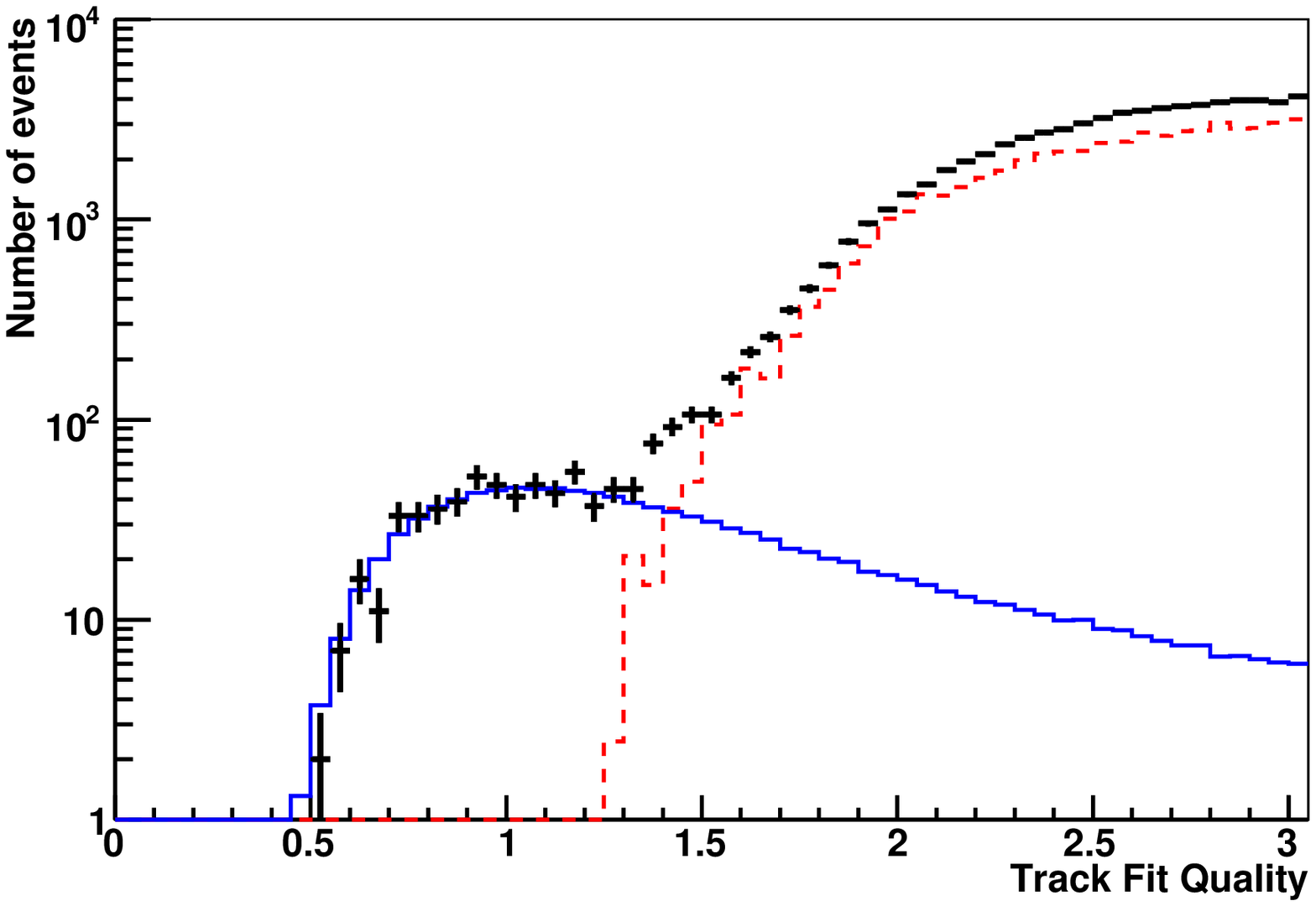}}
      \put(8,0){\epsfbox{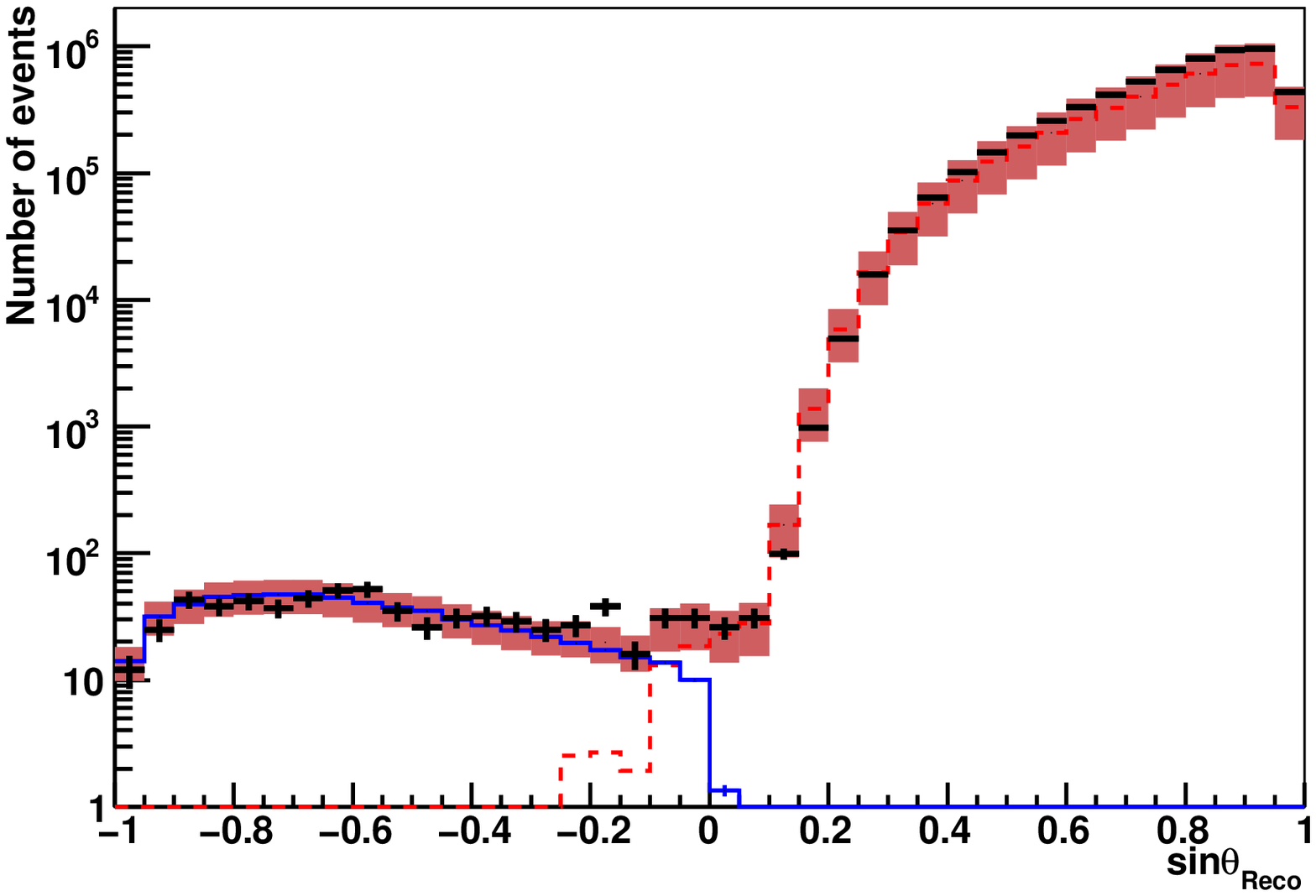}}
    \end{picture}
\caption{{\bf Left}: track fit quality for all upward reconstructed multi-line tracks
for 2008 data (points with error bars) compared to a Monte Carlo sample of downward-going 
atmospheric muons (dashed histogram) and upward-going atmospheric neutrinos (solid histogram); 
{\bf right}: Elevation angle distribution for events with $\tilde{Q}<1.4$. The grey band indicates 
the error band of the combined prediction from neutrino and muon simulations.}
\label{purity}
\end{figure}

Figure~\ref{purity} confronts data and the two Monte Carlo samples for the full 173 days livetime.
The left plot of Figure~\ref{purity} shows the fit quality for the multi-line track fit
of all the tracks reconstructed as upward-going. 
A cut on the track fit quality can
effectively suppress background from misreconstructed atmospheric muons
while maintaining a good efficiency for the atmospheric neutrino sample. A cut in the
track fit quality of $\tilde{Q}<1.4$ ensures a 90\% purity while keeping 48\% of the total sample
of upward reconstructed multi-line neutrino tracks. The cut selects 665 neutrino candidates in data
(about 4 events per day), to be compared to 609 events from the atmopheric neutrino Monte Carlo 
sample and 40 misreconstructed downward-going atmospheric muon events. No event from the coincident
muon sample survives the cut which gives a 90\% upper limit of 13 such events in the final sample.
Data and simulation agree within
the shown statistical errors of the data for $\tilde{Q}<1.4$, whereas for $\tilde{Q}>1.4$ there is a
systematic 30\% excess of data. This excess is well within the estimated systematic error of the
atmospheric muon simulation (see below). 
The resulting elevation angle distribution of the selected events with $\tilde{Q}<1.4$ is shown on 
the right plot of Figure~\ref{purity}. An excellent agreement 
between data and the atmospheric neutrino Monte Carlo
sample based on the Bartol flux~\cite{bartol} is observed for the upward-going events, 
whereas the downward-going part features again a 30\% excess of data with respect to 
the atmospheric muon simulation. The grey band gives the systematic error of the simulation.
An extensive discussion of systematic effects is given in~\cite{atmmuon}. Main contributions are from
the uncertainty in the effective area of the PMTs (10\%), the uncertainty in the water absorption
length (10\%) and the uncertainty in the angular acceptance of the PMTs. The latter is better known
close to the PMT axis (15\% error) than in the backward hemisphere (up to 35\% error) which leads to a
larger total systematic error of 45\% for downward-going atmospheric muons than for upward-going atmospheric
neutrinos (32\%). Within the quoted errors, data are compatible with the chosen flux models both for
atmospheric neutrinos and for atmospheric muons. 

\begin{figure}
\centering
    \begin{picture}(16,6)
      \epsfclipon 
      \epsfxsize=8cm 
      \put(0,0){\epsfbox{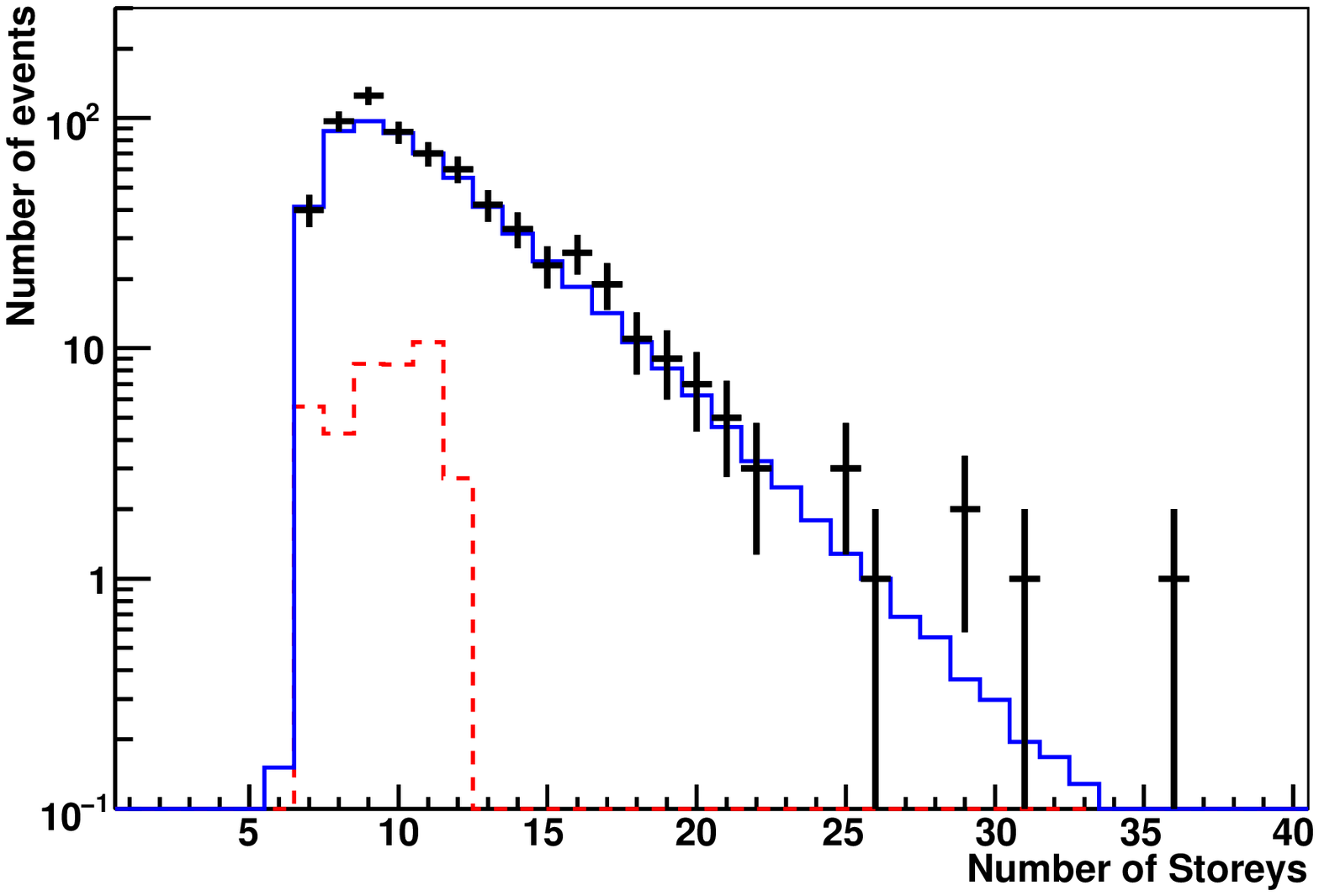}}
      \put(8,0){\epsfbox{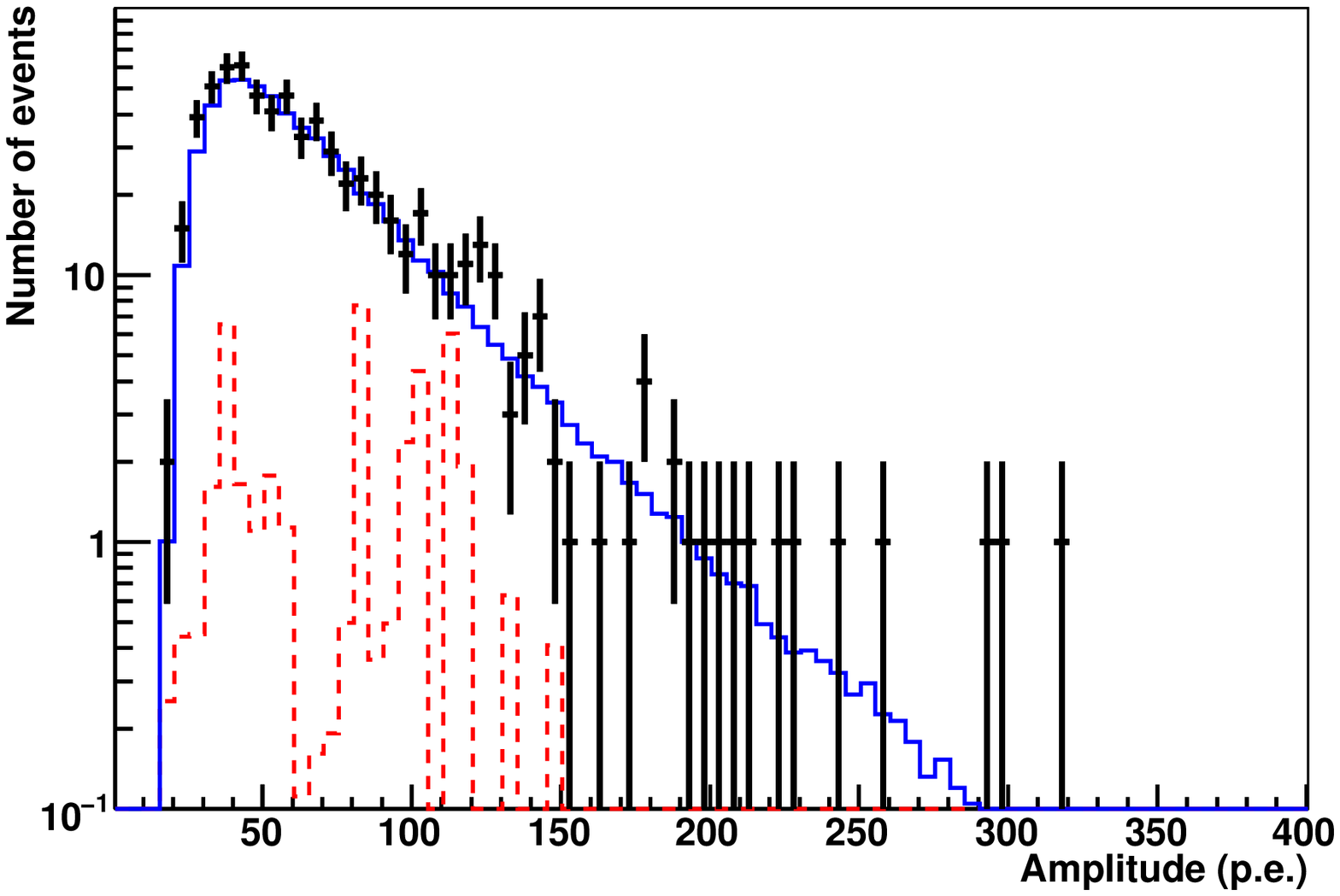}}
    \end{picture}
\caption{{\bf Left}: Number of storeys and {\bf Right}: total charge used in the 
track fit for all upward reconstructed multi-line tracks for events with $\tilde{Q}<1.4$.
Data (points with error bars) are compared to a Monte Carlo sample of downward-going 
atmospheric muons (dashed histogram) and upward-going atmospheric neutrinos (solid histogram).}
\label{nhit-amp}
\end{figure}

Figure~\ref{nhit-amp} compares data and simulation of some related quantities as the number of 
storeys used in the fit and the total charge of all hits used in the fit. Both plots show the
comparison for the selected sample $\tilde{Q}<1.4$, which is dominated by atmospheric neutrinos. 
The shape of both distributions in data is well reproduced in the simulations.

\begin{figure}[bht]
    \begin{picture}(16,6)
      \epsfclipon 
      \epsfxsize=8cm 
      \put(4,0){\epsfbox{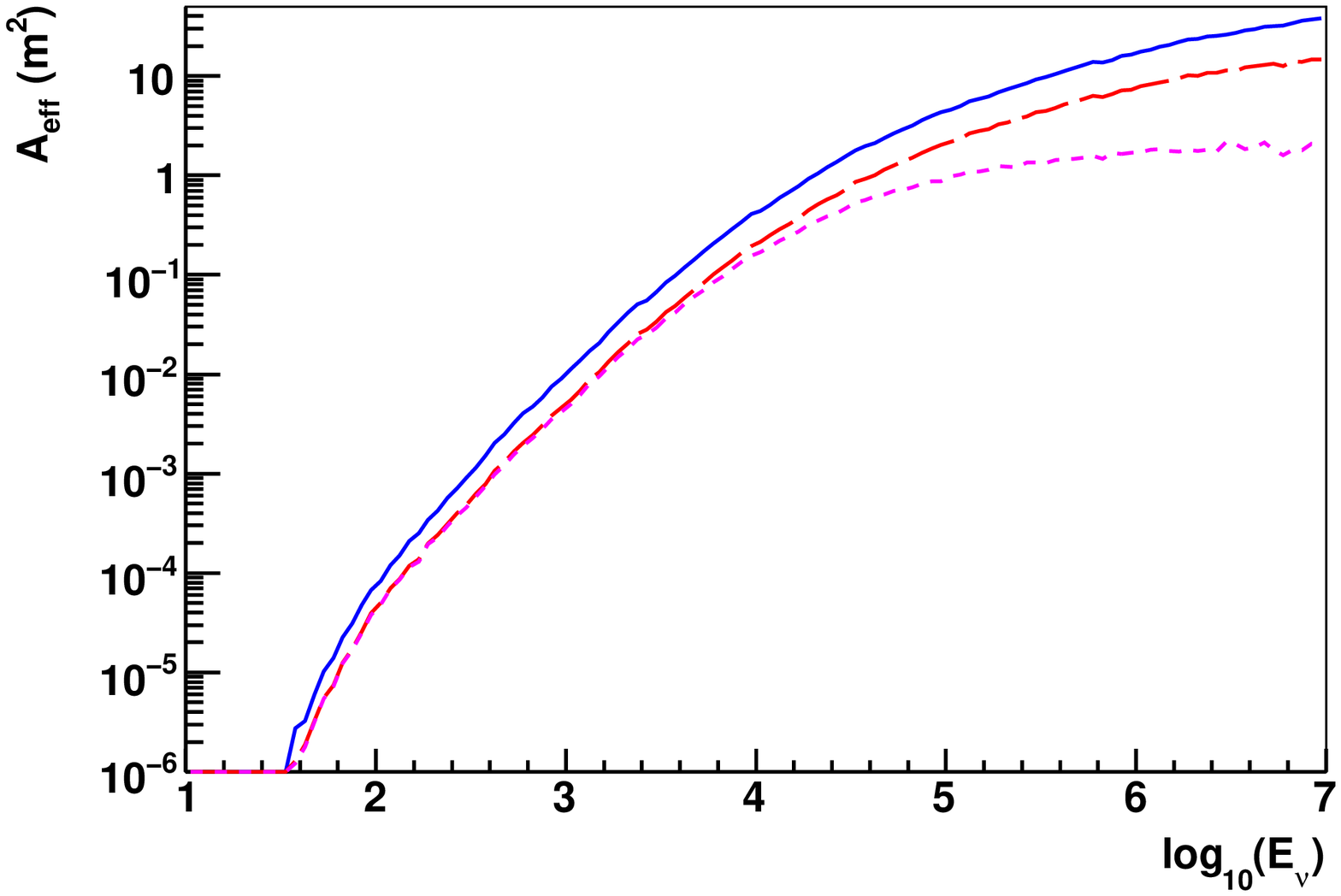}}
    \end{picture}
\caption{Effective area for all multi-line events (solid), selected events with $\tilde{Q}<1.4$ 
(dotted) and selected events with $\tilde{Q}<1.3+(0.04\cdot NDF)^2$ (dashed)
as function of the true neutrino energy for upward-going neutrinos.}
\label{fig:enu}
\end{figure}

Figure~\ref{fig:enu} shows the effective area for all 
reconstructed multi-line events as function of the true neutrino energy. 
The comparison of the solid and the dotted line indicates that
more than 40\% of the events pass the quality cut of $\tilde{Q}<1.4$ 
for energies below 10~TeV. This fraction reduces to 20\% at 100~TeV and 
decreases further for higher energies. The low efficiency at highest energy can be 
explained by the use of a quality cut, which has been derived from Figure~\ref{purity}
for atmospheric neutrinos. By modifying the selection criteria one can easily recover the
high energy part. The condition $\tilde{Q}<1.3+(0.04\cdot NDF)^2$ is used for the dashed line.
It has the same performance as $\tilde{Q}<1.4$ for $E_\nu<10$~TeV where atmospheric neutrinos 
dominate, but a much improved performance at $E_\nu>10$~TeV. $NDF=N_{storey}-5$ is here 
used as a simple estimator for the brightness and therefore energy of an event. This modified cut
selects 624 events in data to be compared to 588 atmopheric neutrinos and 40 atmospheric muons 
keeping the quoted 90\% purity of the neutrino sample. A full likelihood fit~\cite{aart} 
has a similar performance at high energies when requiring a 90\% purity for atmospheric neutrinos. 
For energies below 1~TeV the method presented here is found more efficient than the likelihood fit.

\begin{figure}[bht]
    \begin{picture}(16,6)
      \epsfclipon 
      \epsfxsize=8cm 
      \put(0,0){\epsfbox{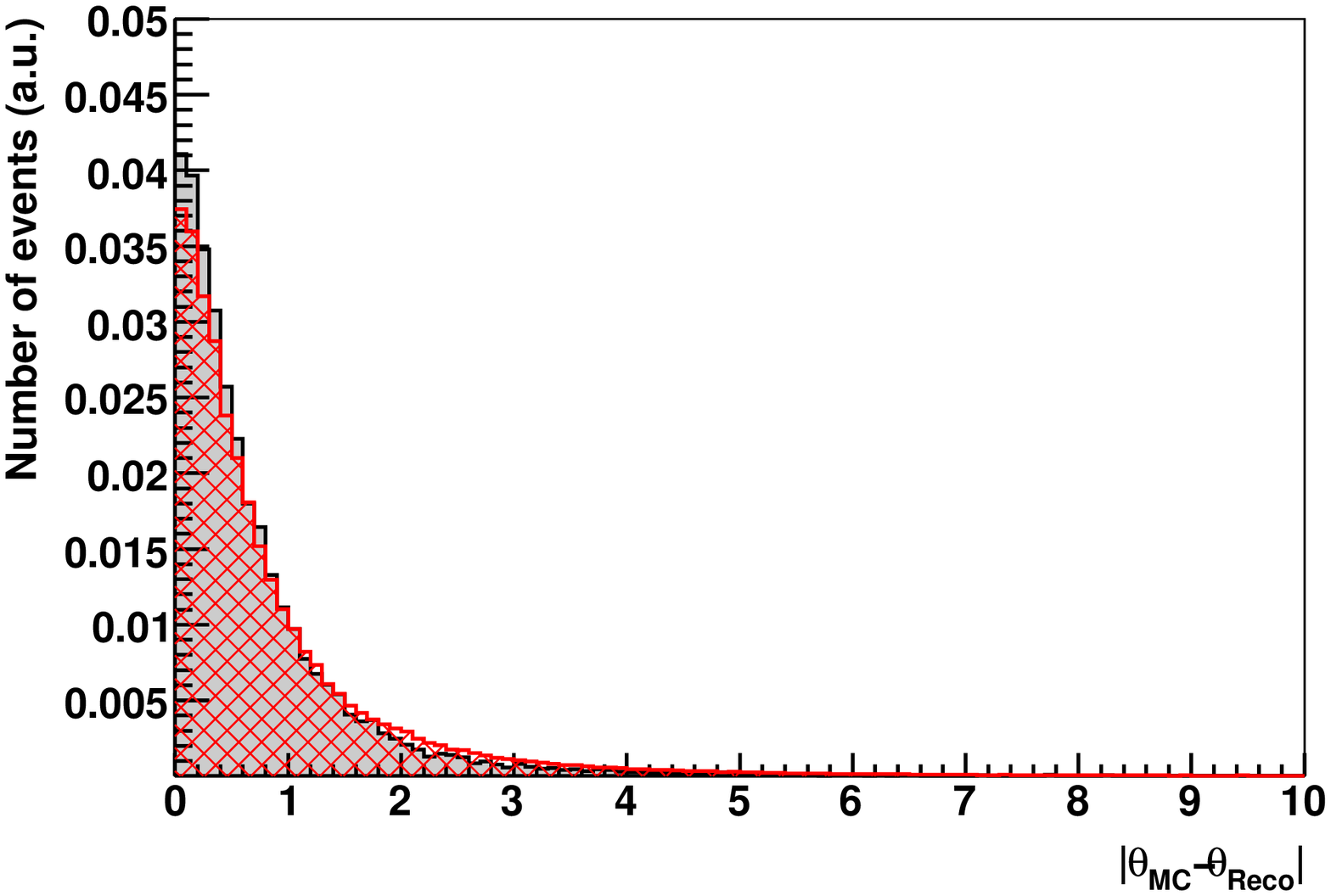}}
      \put( 8,0){\epsfbox{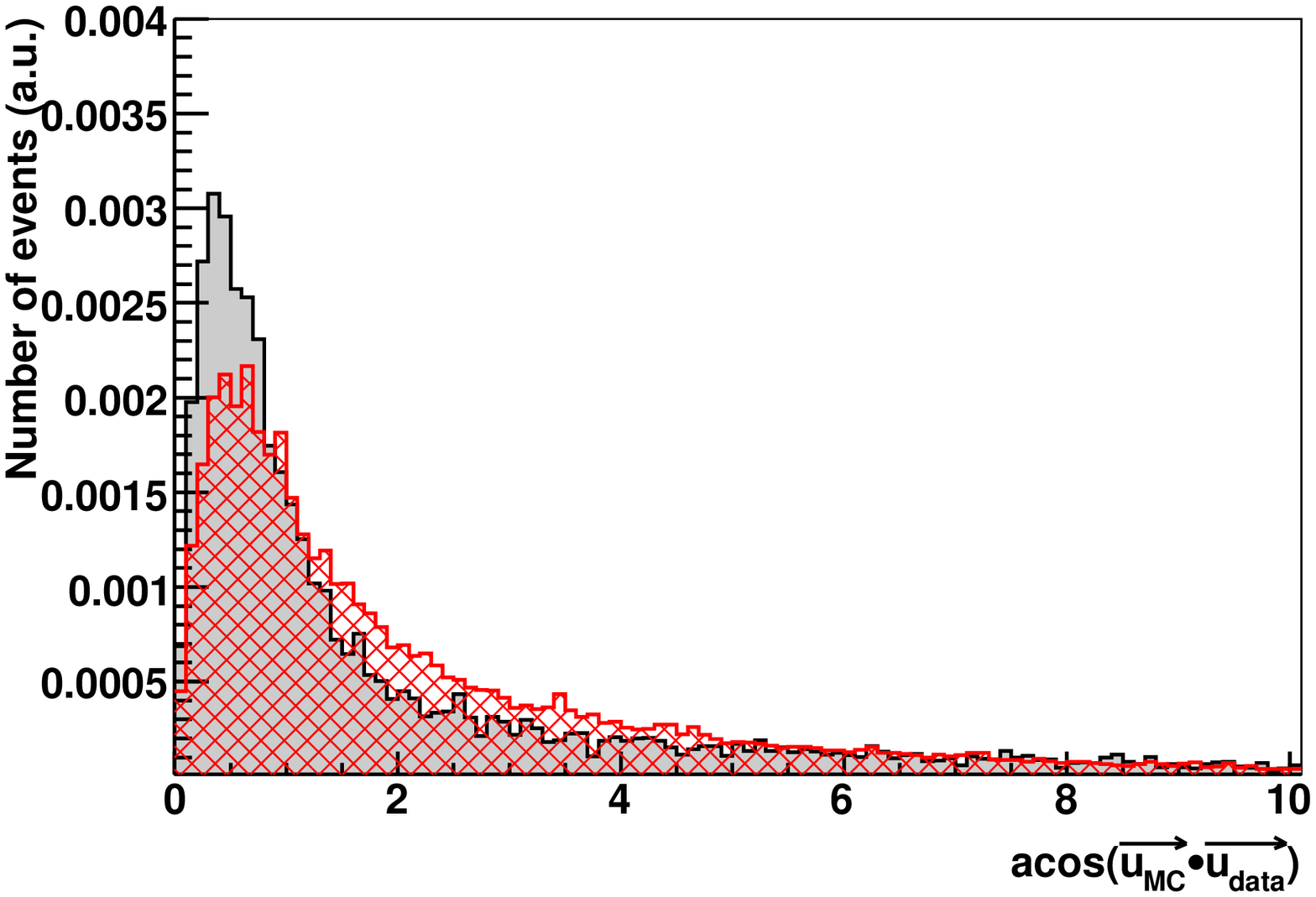}}
    \end{picture}
\caption{The distributions, normalized to the same content, of the 
angular error in degrees for selected atmospheric
neutrinos (full) and atmospheric 
muons (hatched);
{\bf left}: elevation angle error for multi-line events; {\bf right}: space angle error 
for non-coplanar hit selection.}
\label{fig:diffang}
\end{figure}

Figure~\ref{fig:diffang} shows the angular resolution of the data sample which is
selected by a cut of 1.4 on the fit quality.
Both plots of Figure~\ref{fig:diffang} are integrated over the full angular range 
and over the atmospheric neutrino energy spectrum. 
The elevation angle is reconstructed with 
a precision of 0.4$^\circ$  
for upward-going neutrino-induced muons and 0.5$^\circ$ for down-going muons
(median of the angular error compared to the true muon direction). 
The space angle can only be unambiguously defined if the selected hits do not lie
in a plane. 
Mirror solutions in azimuth are otherwise unavoidable.
Because of the simplifications of the detector geometry used during reconstruction, 
a non-coplanar hit selection
can only be achieved when hits from at least three detector lines are used. 
This condition is fulfilled for 25\% of the selected multi-line atmospheric neutrino events.
For this sub-sample a median space angle error of 1.0$^\circ$ is found 
for atmospheric neutrinos and 1.4$^\circ$
for atmospheric muons as shown on the right plot of Figure~\ref{fig:diffang}. 

\subsection{Improving the angular resolution}

The angular resolution which has been quoted in the previous section can be improved
in several ways, without changing the event selection and without 
compromizing the speed of the algorithm.
 
\subsubsection*{Resolving azimuthal degeneracy}

For a particle trajectory reconstructed using hits on only two straight detector lines, 
there always exists an alternative trajectory
having an identical $\tilde{Q}$-value, but a different direction. The two
trajectories will have the same elevations, but will differ in
their azimuthal orientation. The degenerate trajectory is easily
determined given the positions of the two detector lines and
the trajectory resulting from the fit. First, the point at which the
particle trajectory crosses the plane connecting the two detector lines
is determined. The trajectory is then rotated about this point by an
angle $2\psi$ to form the degenerate track candidate, where $\psi$ is the
angle between the fit result and the plane connecting the two detector
lines.


Armed with both possible track solutions, the two candidates can be 
discriminated with the following procedure.
The time residuals of all hits are calculated from the differences of the
actual hit times and the theoretical hit times according to Equation~\ref{thit}. 
The sum
of the charge of each hit having a
time residual smaller than 30~ns is then calculated. This can
include hits from any line and not just the two used in the original
fit, helping to break the degeneracy. The hit selection is done
separately for both track candidates and the candidate with the largest
total charge is chosen. In simulations of both
downward-going muons and upward-going neutrinos this procedure chooses the 
correct track candidate 70\% of the time and improves the two-line angular resolution by
42\% for upward-going atmospheric neutrinos and by
15\% for downward-going atmospheric muons. 

\subsubsection*{Additional fit step}

Taking the result of a multi-line track fit or its mirror solution 
as it was described in the previous sections as a prefit, 
a new hit selection and a subsequent fit is performed. 
Every hit which has a time residual smaller than 20~ns
with respect to the prefit track is chosen.  These hits are then used to minimize the function
\begin{equation}
M = \sum^{N_{hit}}_{i=1}\left[2 \sqrt{1 + \frac{(t_{\gamma} - t_i)^2}{2\sigma^2}} - 2\right],
\label{Mchi2}
\end{equation}
which is a robust estimator that combines the properties of both 
least-squares and absolute-value minimizers~\cite{zhang1997}. This estimator
is quadratic for small values of the time residual ({\it i.e.} when
$|t_{\gamma} - t_i|/\sigma \ll 1$), just as with a standard $\chi^2$
estimator, but is linear for large time residuals. It is therefore not
affected as dramatically by background hits with large residuals
surviving the hit selection. Note that only the time residual of hits
are minimized at this stage; no information concerning the number of
photoelectrons detected is used. For the current studies $\sigma = 1$~ns has been taken. 
The precise value of $\sigma$ has little impact on the
angular resolution.

\subsubsection*{Use of the detailed detector geometry}

While the algorithm has so far assumed a simplified detector geometry,
it is possible to extend it in order to
exploit the actual location and orientation of the PMTs.
This detailed geometry information can be employed in the calculation 
of the predicted photon arrival time ($t_\gamma$ in Equation~\ref{Mchi2}). 
The use of the resulting time residual for both the hit 
selection and track fit minimization, as described in the previous section, 
has been studied.  Simulations for this study have been performed in the absence
of sea currents, {\it i.e.} again for straight detector lines.
The systematic angular shift due to the coherent inclination of all
detector lines in a horizontal sea current is estimated to be smaller than $0.2^\circ$ for a 
typical sea current of 5~cm/s as measured at the ANTARES site. 
However, such a systematic shift is not included in the numbers and figures quoted below.

\subsubsection*{Results}

The effect of these three additional steps can be seen in
Table~\ref{tab:fullfitresults}, which compares the median angular error
of the multi-line algorithm described in Section~\ref{sec:trackfit} to
the subsequent fit using Equation~\ref{Mchi2} with and without the use of a
detailed detector geometry. The condition on the number of detector lines $N$, 
from which hits have been used in the fit, 
is specified in the first column of Table~\ref{tab:fullfitresults}.
Results are given for upward-going
atmospheric neutrinos as well as for neutrinos from an $E_\nu^{-2}$ flux, 
which is typical for astrophysical sources. Only events which fulfill the condition 
$\tilde{Q}<1.3+(0.04\cdot NDF)^2$ are included.
 
For tracks reconstructed using hits on only two lines in the initial fit,
the improvement can be attributed mainly to
the partial resolution of the azimuthal symmetry. 
For the case where more than two lines had been used initially, 
the main improvement comes from the additional fit step.
An additional $0.1^\circ$ in resolution may be gained by using
the detailed detector geometry within each
storey. The full multi-line sample ($N>1$) profits from a combination of both effects.

\begin{table}[htpb]
\begin{center}
\begin{tabular}{||c||c|c|c||c|c|c||} \hline
      & \multicolumn{3}{c||}{$\tilde{\alpha}$ for atmospheric $\nu$} 
      & \multicolumn{3}{c||}{$\tilde{\alpha}$ for $\Phi_\nu\propto E^{-2}$} \\ \hline
lines & Eq.~\ref{chi2} & Eq.~\ref{Mchi2} & Eq.~\ref{Mchi2} (geom) 
& Eq.~\ref{chi2} & Eq.~\ref{Mchi2} & Eq.~\ref{Mchi2} (geom) \\ \hline 
N=2 & 4.2$^{\circ}$ &1.7$^{\circ}$ &  1.5$^{\circ}$  & 4.4$^{\circ}$ & 1.3$^{\circ}$& 1.1$^{\circ}$  \\
N$>$1 & 2.9$^{\circ}$ &1.3$^{\circ}$ & 1.1$^{\circ}$  & 1.3$^{\circ}$& 0.60$^{\circ}$ & 0.52$^{\circ}$  \\
N$>$2 & 1.00$^{\circ}$ &0.64$^{\circ}$ & 0.54$^{\circ}$  & 0.94$^{\circ}$ & 0.47$^{\circ}$& 0.41$^{\circ}$  \\ \hline
\end{tabular}
\end{center}
\caption{The median angular error ($\tilde{\alpha}$) in the multi-line fit
algorithm as of Section~\ref{sec:trackfit} and the subsequent fit with and without using
geometry information, for atmospheric neutrinos and neutrinos from an $E^{-2}$ flux.}
\label{tab:fullfitresults}
\end{table}

\begin{figure}[bht]
    \begin{picture}(16,6)
      \epsfclipon 
      \epsfxsize=8cm 
      \put(0,0){\epsfbox{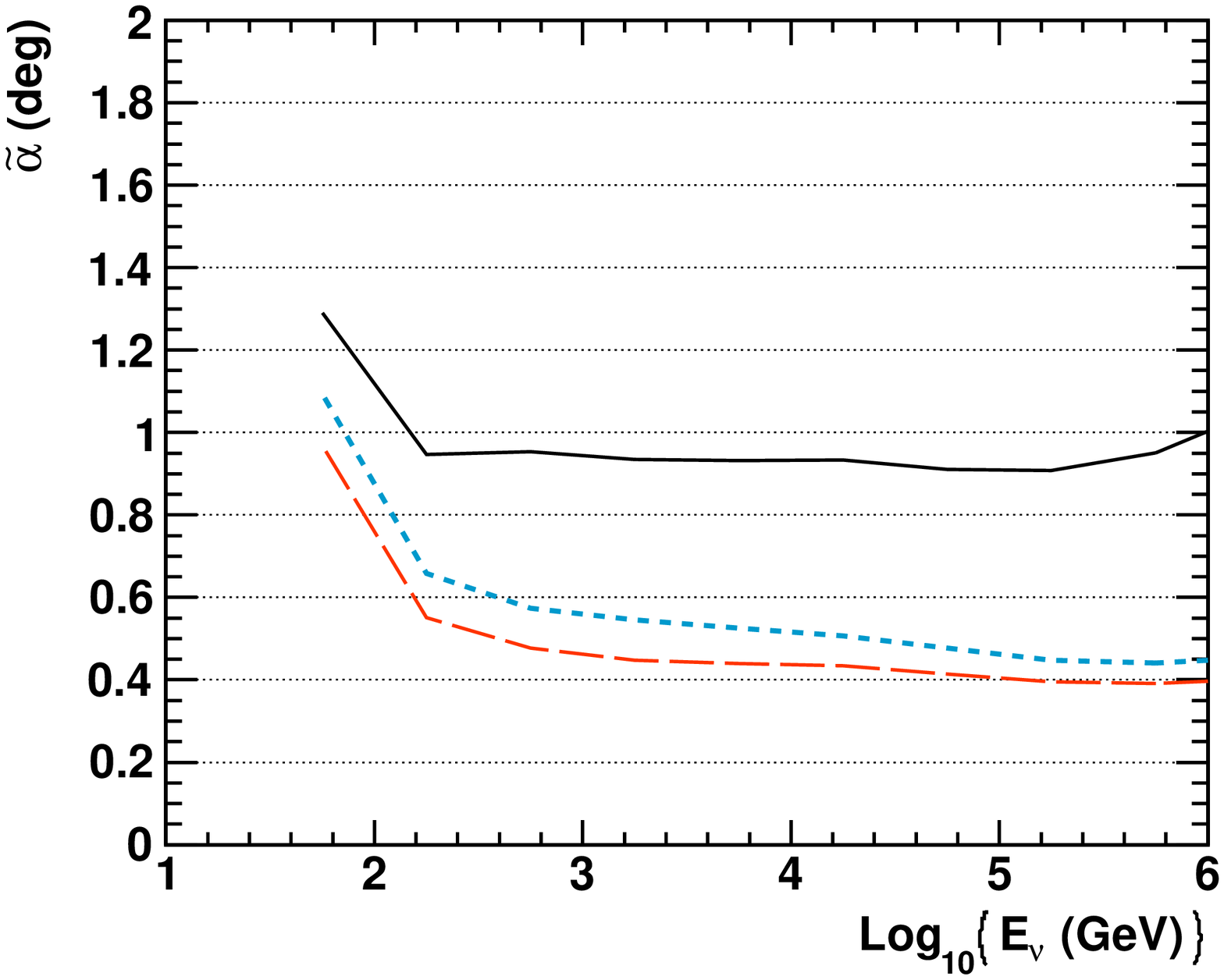}}
      \put(8,0){\epsfbox{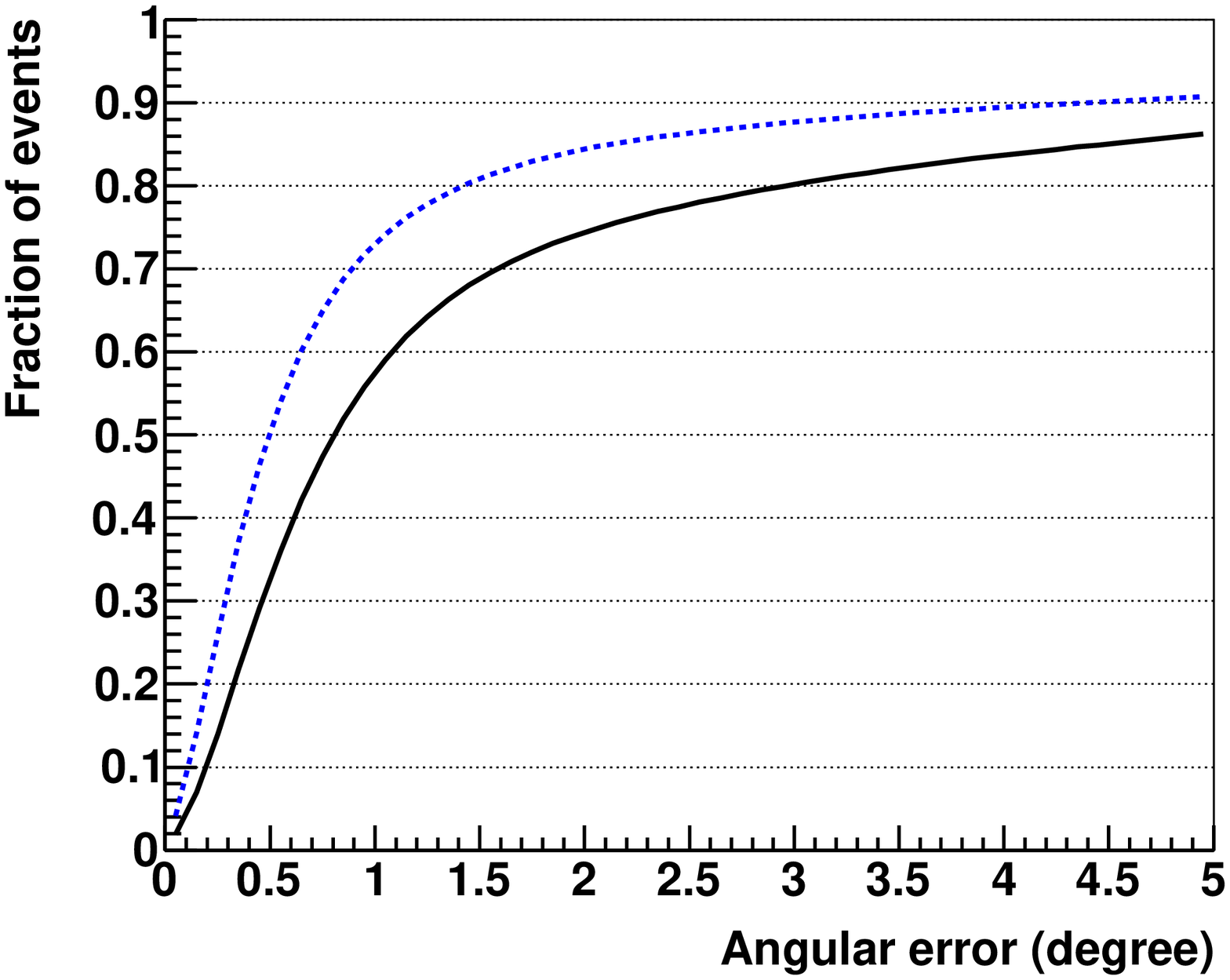}}
    \end{picture}
\caption{{\bf Left}: Dependence of the angular resolution for selected tracks 
($\tilde{Q}<1.3+(0.04\cdot NDF)^2, N>2$) on $E_\nu$. The median angular error of the multi-line fit (solid line) is
compared against the fit from Equation~\ref{Mchi2} without (dotted) and with (dashed)
use of a detailed description of the geometry of a storey;
{\bf right}: Cumulative point spread function for selected upward-going neutrino events 
with an $E^{-2}$ spectrum and $\tilde{Q}<1.3+(0.04\cdot NDF)^2, N>2$ for multi-line fits (solid) and fits from
Equation~\ref{Mchi2} with an approximate geometry (dotted).}
\label{fit:fullfitres}
\end{figure}

The
improvement in the angular error for neutrino-induced muons can also be
seen in Figure~\ref{fit:fullfitres} (left), which shows the energy dependence of the median
angular deviation from the true muon direction for all reconstructed tracks
using hits on more than 2 lines having a $\tilde{Q}$-value better
than $1.3+(0.04\cdot NDF)^2$. The selection is applied only to the original track fit, 
so that the observed improvement of the angular resolution is
the result a refined hit selection, combined with a new fit. 
The three curves correspond to the multi-line track fit 
from Section~\ref{sec:trackfit} (solid) and the fits using Equation~\ref{Mchi2} 
with an approximated geometry (dotted) and with a detailed geometry 
of each storey (dashed). The muon track angular
error is almost energy independent at neutrino energies above 10~TeV.
It converges to the values quoted in table~\ref{tab:fullfitresults} for an $E^{-2}$ flux.

This can be compared to the performance of a full likelihood fit~\cite{aart} 
which uses the full geometry, a detailed modelling of the time residual, 
({\it e.g.} an energy dependent tail of late hits and a contribution
of noise hits) and which applies a scan of starting positions of the fit to avoid local minima. 
Such a procedure can reach an angular resolution of $0.3^\circ$ at high energies for the entire
multi-line sample, while using about 10 times more computing time than the method presented here. 

Figure~\ref{fit:fullfitres} (right) shows the point spread function for neutrino events after
the fit from Equation~\ref{Mchi2}. Contrary to the figures shown earlier the angular error 
is given here with respect to the neutrino direction  - a quantity which is relevant to point back to
a potential neutrino source. The optical followup program~\cite{tatoo} uses a sub-sample of the 
events, shown on the dashed line in Figure~\ref{fit:fullfitres} (right). 
As the field of view of the used robotic telescope is of the order
of $2^\circ$, it can be seen that more than 70\% of the candidate sources would lay inside the field of view.

\begin{figure}[bht]
    \centering
    \begin{picture}(8,6)
      \epsfclipon 
      \epsfxsize=8cm 
      \put(0,0){\epsfbox{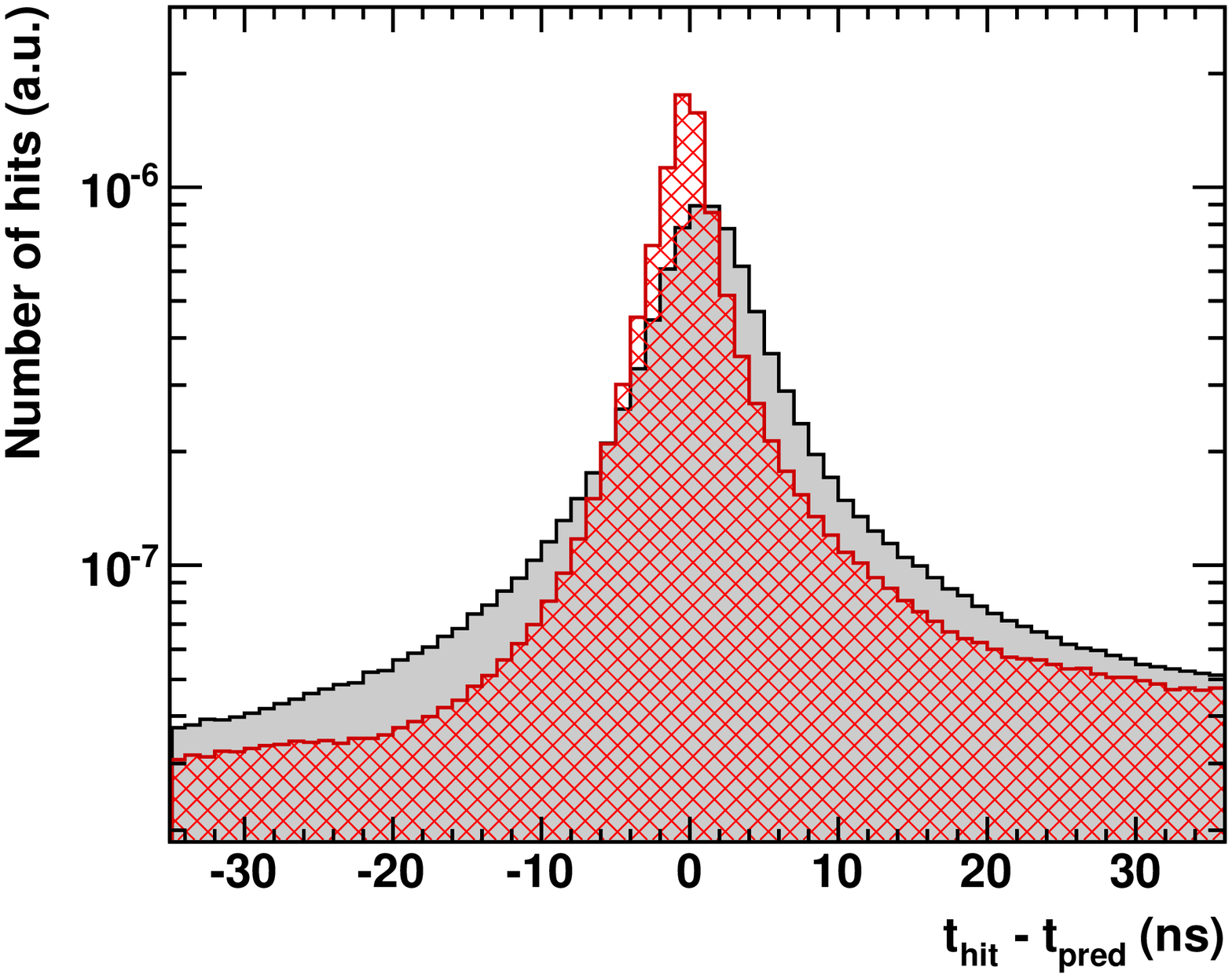}}
    \end{picture}
\caption{Time residuals of all hits (not just those used in the fit) for the
multi-line track fit (full) and the
subsequent fit using geometry information (hatched).} 
\label{fit:timres}
\end{figure}

The improvement is further demonstrated by the time residual
distributions, as shown in Figure~\ref{fit:timres}. The time
residuals are shown for all hits in each event for atmospheric neutrino events.
The residuals of the subsequent fit using the detailed geometry
show a more pronounced peak at zero, whose width is roughly 40\% smaller
compared to the multi-line fit from Section~\ref{sec:trackfit}.

These extensions to the algorithm do not significantly affect the time
needed for track reconstruction.
Including the subsequent fit, the entire reconstruction algorithm  takes
an average of 5~ms per muon event on an Intel Xeon 2.66~GHz CPU
for events recorded by the 12-line ANTARES detector.

\section{Conclusions}

A generic and fast track fit algorithm for muon tracks in neutrino telescopes 
has been presented which can reliably distinguish 
upward-going neutrinos from the overwhelming background of downward-going muons. 
The method is used in ANTARES
for different
applications such as an online event display, a neutrino monitor and an alert 
sending program to trigger
optical follow-ups of selected neutrino events. It has also been used for an analysis of 
atmospheric muons~\cite{atmmuon}. Other potential applications include the study of
magnetic monopoles and nuclearites and the measurement of atmospheric neutrinos.

\section{Acknowledgement}

The authors acknowledge the financial support of the funding agencies: 
Centre National de la Recherche Scientifique (CNRS), 
Commissariat \`{a} l'Energie Atomique et aux Energies Alternatives  (CEA), 
Agence National de la Recherche (ANR), 
Commission Europenne (FEDER fund and Marie Curie Program), 
R\'{e}gion Alsace (contrat CPER), 
R\'{e}gion Provence-Alpes-Cte d'Azur, 
D\'{e}partement du Var and Ville de La Seyne-sur-Mer, France; 
Bundesministerium f\"{u}r Bildung und Forschung (BMBF), Germany; 
Istituto Nazionale di Fisica Nucleare (INFN), Italy; 
Stichting voor Fundamenteel Onderzoek der Materie (FOM), 
Nederlandse organisatie voor Wetenschappelijk Onderzoek (NWO), the Netherlands; 
Council of the President of the Russian Federation for young scientists 
and leading scientific schools supporting grants, Russia; 
National Authority for Scientific Research (ANCS), Romania; 
Ministerio de Ciencia e Innovaci\'{o}n (MICINN), 
Prometeo of Generalitat Valenciana and MultiDark, Spain. 
Technical support of Ifremer, AIM and Foselev Marine 
for the sea operation and the CC-IN2P3 for the computing facilities is acknowledged.


\end{document}